\documentclass[fleqn,usenatbib]{mnras}

\usepackage{newtxtext,newtxmath}

\usepackage[T1]{fontenc}
\usepackage{ae,aecompl}

\usepackage{graphicx}	
\usepackage{amsmath}	
\usepackage{nicefrac,xfrac}
\usepackage{verbatim}
\usepackage{bbold}

\usepackage[rgb]{xcolor}
\usepackage{smartdiagram}
\usetikzlibrary{shapes.symbols}
\usepackage{pdflscape}

\usepackage{color,ulem}

\usepackage{mdframed}
\usepackage[export]{adjustbox}

\newcommand{\be}{\begin{equation}}
\newcommand{\ee}{\end{equation}}
\newcommand{\ba}{\begin{eqnarray}}
\newcommand{\ea}{\end{eqnarray}}
\newcommand{\mperh}{\,h^{-1}\,{\rm Mpc}}

\newcommand{\mbi}[1]{\mbox{\boldmath$#1$}}

\title[BIRTH of the COSMOS Field]{BIRTH of the COSMOS Field: Primordial and Evolved Density Reconstructions During Cosmic High Noon}

\author[Metin Ata et al.]{
Metin Ata$^{1}$\thanks{E-mail: metin.ata@ipmu.jp, Kavli IPMU Fellow},
Francisco-Shu Kitaura$^{2,3}$, 
Khee-Gan Lee$^{1}$,
Brian C. Lemaux$^{4}$,
\newauthor{ Daichi Kashino$^{5}$, Olga Cucciati$^{6}$,  M{\'o}nica Hern{\'a}ndez-S{\'a}nchez$^{2,3}$  \&  Oliver Le F{\`e}vre$^{7}$}
\\
$^{1}$Kavli Institute for the Physics and Mathematics of the Universe (Kavli IPMU), WPI,\\  The University of Tokyo Institutes for Advanced Study (UTIAS), The University of Tokyo, Kashiwa, Chiba, 277-8568, Japan\\
$^{2}$Instituto de Astrof\'{\i}sica de Canarias (IAC), Calle V\'{\i}a Lactea s/n, 38200, La Laguna, Tenerife, Spain \\ 
$^{3}$Departamento de Astrof\'{\i}sica, Universidad de La Laguna (ULL), E-38206, La Laguna, Tenerife, Spain\\ 
$^{4}$Department of Physics, University of California, Davis, One Shields Ave., Davis, CA 95616, USA \\
$^{5}$Department of Physics, ETH Z{\"u}rich, Wolfgang-Pauli-Strasse 27, CH-8093 Z{\"u}rich, Switzerland\\
$^{6}$INAF - Osservatorio di Astrofisica e Scienza dello Spazio di
Bologna, via Gobetti 93/3, 40129 Bologna, Italy\\
$^{7}$Aix Marseille Universit\'e, CNRS, LAM (Laboratoire
d'Astrophysique de Marseille) UMR 7326, 13388 Marseille, France 
}

\date{Accepted XXX. Received YYY; in original form ZZZ}

\pubyear{2020}

\begin{document}
\label{firstpage}
\pagerange{\pageref{firstpage}--\pageref{lastpage}}
\maketitle

\begin{abstract}
This work presents the first comprehensive study of structure formation at the peak epoch of cosmic star formation over $1.4\leq z \leq 3.6$ in the COSMOS field, including the most massive high redshift galaxy proto-clusters at that era. 
We apply the extended \texttt{COSMIC BIRTH} algorithm to account for a multi-tracer and multi-survey Bayesian analysis at Lagrangian initial cosmic times. Combining the data of five different spectroscopic redshift surveys (zCOSMOS-deep, VUDS, MOSDEF, ZFIRE, and FMOS-COSMOS), 
we show  that the corresponding unbiased primordial density fields can be inferred, if a proper survey completeness computation from the parent photometric catalogs, and a precise treatment of the non-linear and non-local evolution on the light-cone is taken into account, including (i) gravitational matter displacements, (ii) peculiar velocities, and (iii) galaxy bias. 
The reconstructions reveal a holistic view on the known proto-clusters in the COSMOS field and the growth of the cosmic web towards lower redshifts. The inferred distant dark matter density fields concurrently with other probes like tomographic reconstructions of the intergalactic medium will explore the interplay of gas and dark matter and are ideally suited to study structure formation at high redshifts in the light of upcoming deep surveys. 
\end{abstract}




\section{Introduction}
\label{sec:intro}
Our current standard cosmological model predicts a hierarchical clustering of subsequently merging small structures to greater ones \citep[see][]{White1978,Fry1978} up to the formation of the largest galaxy super-clusters observed in our present local Universe \citep[e.g.][with a total mass of $ \sim 10^{17} M_{\sun}$]{Laniakea}.
The formation history and the exploration of the underlying physical phenomena of galaxy clustering over cosmic history remains an important question \citep[see the pioneering work of][and references therein]{Kauffmann1998}. 

Furthermore, progenitors of galaxy clusters and super-clusters and their halos are key probes to understand early structure formation \citep{Cohn2008, Gao2018} and can be used to constrain a particular dark matter model \citep{2001ApJ...556...93B}.
Also, the analysis of galaxy clusters allow us to study galaxy formation, test cosmological parameters \citep{Allen2011} and constrain non-standard cosmological models \citep[][]{Kravtsov2012,CostanziAlunnoCerbolini2013}.

In particular, the range of $2 \lesssim z \lesssim 3$ marks the peak epoch of star formation in the Universe, frequently referred to as ``Cosmic High Noon'' \citep{starformation}. The processes driving star formation of galaxies and clusters have been studied in previous works, showing a nontrivial relation of star formation and the galaxies' environmental density \citep{Cooper2008,Koyama2013,Kawinwanichakij2017,Muldrew2018,Ji2018}.
Moreover, the quenching of star formation within massive galaxy clusters has been found \citep{Cooper2008}, suggesting on average a passive evolution of galaxy clusters during the last $\approx 10$ Gyr and studying an environmental dependency \citep[see][]{Lemaux2012,Belfiore2017,Tomczak2019,Lemaux2019}.
Therefore, an accurate description of the dark matter density distribution at these redshifts will potentially help to understand the interconnection of  star formation \citep[see e.g.][]{2013ApJ...770...57B, Madau2014,Shi2019} and the location within the cosmic web \citep[e.g.][]{1996Natur.380..603B}. Detailed analyses of high redshift structures have been accessible via numerical simulations \citep[e.g.][]{Watson2013,Henriques2014,Chiang2017}, studying their formation history and the importance of the environment. 

While wide field galaxy surveys like SDSS-III Baryon Oscillation Spectroscopic Survey (BOSS) \citep{BOSS2016} and 2MASS \citep{Skrutskie2006} have played an important role in spatially mapping the large-scale structure of galaxies and quasars \citep[see e.g.][]{Ata2018}, these have focused on scales of the baryon acoustic oscillations (BAOs) \citep{Eisenstein2005,Percival2007}.
Some endeavours have been done to map the structure formation down to megaparcec scales over to smaller footprints in so-called pencil-beams\footnote{Pencil-beam surveys do not statistically resolve the transverse large-scale structures at $\sim 10~\mperh$ scales.},
such as ALMA deep fields \citep{Casey2018} reaching out to $z\sim1.5-2.5$. However, no spectroscopic surveys have abundantly mapped the large-scale density distribution down to megaparcec scales beyond $z>1.5$, although multiple projects aim to push beyond this boundary in the near future, such as DESI \citep{Levi2013}, EUCLID \citep{Amendola2018}, PFS \citep{10.1093/pasj/pst019}, 4MOST \citep{DeJong2012}, and MOONS \citep{Cirasuolo2014}.
For this reason we are still missing observational validation of quasi-linear structure formation at about $z\gtrsim 1.5$, before non-linearities start dominating the emerging cosmic web.

Over nearly two decades, the Cosmic Evolution Survey (COSMOS)  \citep[see latest version][]{cosmos2015} \citep{Capak2007,Scoville2007} has been a major ongoing effort to photometrically observe galaxies across a sufficiently wide footprint to resolve transverse large-scale structure, while simultaneously having sufficient depth to probe the entire span of cosmic history from the Local Universe right into (and eventually beyond) the Epoch of Reionization \citep[e.g.][]{Scoville2013}. Observations within COSMOS have identified large aggregations of galaxies at high redshifts \citep{cluster1}, i.e.\ galaxy proto-clusters, that have been extensively studied over a large redshift range at Cosmic Noon  \citep{Diener2013, Diener2015, Chiang2015, Casey2015,Wang2016, Lee2016,Cucciati2018,Lemaux2018,Darvish2020}. These overdense structures have been proposed to be possible progenitors of the Coma-like galaxy clusters, potentially assuming total masses of $\sim (1-2) \times 10^{15}M_{\sun}$ at $z=0$  \citep{Lee2016,Cucciati2018,Lemaux2018,Darvish2020}. The COSMOS field is therefore ideally suited to study early structure formation and the evolution of galaxy proto-clusters into the mature structures we observe at the current epoch.

However, some of the aforementioned pioneering studies have typically been based on individual spectroscopic surveys in the COSMOS field and also not taking selection criteria into account. Therefore, a consistent analysis of the confirmed structures combining the multiple deep galaxy spectroscopic surveys over a sufficiently large redshift range is still missing. This has led to a heterogeneous view of structures that may or may not co-evolve, e.g. the reported overdensities at $z\approx 2.4-2.5$ that are within $\sim 100 \mperh$ of each other \citep{Diener2015, Chiang2015, Casey2015, Wang2016}. Moreover, analysis of the overdensities have typically been carried out by comparing with analogous structures in $N$-body simulations rather than direct analysis of the observed structures, which leads to greater uncertainties due to the diversity of structures with similar aggregate properties in the simulations (e.g.\ aperture mass, 
velocity dispersion etc). This situation compounds an attempt to compare the results of these findings.

Our goal in this work is to recover the dark matter density distribution in the COSMOS field during
Cosmic Noon ($1.4\leq z \leq 3.6$), jointly constrained from different spectroscopic galaxy surveys, revealing a consistent reconstruction of all structures within the field. We infer the Gaussian density field at redshift $z=100$ and the density field corresponding to the observed redshift over the range of $1.4\leq z \leq 3.6$, which we refer to as `initial' and `final' conditions, respectively. Over the last two decades, several density reconstruction methods have been proposed in literature, starting from iterative methods \citep{Zaroubi1995,Wang2009,Kitaura2009}, over non-linear ones using, e.g., a lognormal prior and a Poisson likelihood \citep{Kitaura2010}, to more realistic structure formation models, such as Lagrangian perturbation theory (LPT) \citep{kitaura_kigen,Jasche2013,Wang2013,Schmittfull2017, Hada2018,PatrickBos2019,birth,2019arXiv190906396L} or Particle-Mesh based codes \citep{Wang2014,Jasche2019}. Recent reconstruction approaches also aim to infer initial and final conditions from the observations of absorption lines from the Intergalactic medium \citep[see e.g.][]{2012MNRAS.420...61K,Horowitz2019,Porqueres2019}. 

In this work we use five spectroscopic galaxy surveys in the COSMOS field, (i) zCOSMOS-deep \citep{Lilly2009}, (ii) VIMOS Ultra Deep Survey \citep[VUDS, ][]{LeFevre2015}, (iii) MOSFIRE Deep Evolution Field \citep[MOSDEF, ][]{Kriek2015}, (iv) KECK/MOSFIRE Spectroscopic Survey of Galaxies in Rich Environments \citep[ZFIRE, ][]{ZFIRE2016}, and the (v) FMOS-COSMOS survey \citep{Silverman2015, FMOS} to jointly reconstruct the initial dark matter density field with the \texttt{COSMIC BIRTH} method \citep{birth}, which is ideally suited for this purpose as it deals for the first time with light-cone evolution effects beyond the Zel'dovich approximation. Considering these surveys we firstly develop a formalism to combine them within our Bayesian reconstruction framework. Secondly, we calculate the selection functions of each survey to estimate the completeness of the observations as a function of the angular and radial dimensions.
Thirdly, we study the galaxy bias beyond passive evolution as a function of redshift.

The inferred density fields are ideally suited to compare with the intergalactic medium (IGM) absorption maps, from the $z\sim 2-2.5$ CLAMATO (COSMOS Ly$\alpha$ Mapping And Tomographic Observations) survey  \citep{Lee2014,Lee2014a,Lee2016,Lee2018}, and the $z\sim 2.2-2.8$ LATIS (Ly$\alpha$ Tomography IMACS Survey) \citep{2020ApJ...891..147N}, based on the Lyman-$\alpha$ forest tomography technique \citep{Pichon2001} and thus will provide insights into the relationship of matter clustering and the properties of the IGM. Other future applications include a direct study of galaxy properties as a function of underlying matter density and a full constrained $N$-Body simulation starting from the inferred initial conditions.

This article is structured as follows. In Section \S \ref{sec:birth} we give a brief introduction to the recently developed initial density perturbations reconstructions algorithm \texttt{COSMIC BIRTH}. We discuss the challenges to reconstruct the density perturbations in the COSMOS field, and present the necessary extensions to our algorithm. In Section \S \ref{sec:surveys} we give details of the surveys used in this work, how we selected the data and how we constructed the selection functions. In Section \S \ref{sec:lsbias} we give details of our large-scale bias computation and how we determined the bias evolution with increasing redshift. In Section \S \ref{sec:reconstructions} we describe the application of \texttt{COSMIC BIRTH} to the surveys in the COSMOS field, present the inferred density fields and provide additional diagnostics to validate the \texttt{COSMIC BIRTH} reconstruction algorithm. Finally, we give a summary in Section \S \ref{sec:conc}, discuss our results and forecast future efforts based on this work.  

Throughout this paper we use a fiducial flat $\Lambda$CDM cosmology with a set of cosmological parameters \{$p_{\rm c}$\} of
\ba
\label{eq:cosm}
\{p_{\rm c}\}=\{\Omega_{\rm M}= 0.31, \Omega_{\Lambda}= 0.69, \sigma_8  = 0.82, n_s     = 0.96, h=0.68 \} \,
\ea
concertedly chosen with the parameters in \citet[][]{Lee2018}. 
All distances are given in comoving $\mperh$ units. 

\section{\texttt{COSMIC BIRTH}-Algorithm}
\label{sec:birth} 

Density field reconstructions from observed galaxy distributions are an ongoing effort in cosmological science. The aim of this work is to infer the initial density fluctuations $\delta(\mbi q)$ at Lagrangian coordinated $\mbi q$ given galaxy positions that have been observed in Eulerian redshift-space $\mbi s$.
These two frames can be connected if the peculiar velocity $\mbi v_{\rm r} (\mbi q)$ and displacement $\mbi \psi(\mbi q)$ fields are known, as
\ba
\label{eq:phase}
\mbi q = \mbi s  - \mbi v_{\rm r} (\mbi q)  - \mbi \psi(\mbi q)\, .
\ea
Equation \ref{eq:phase} shows that the mapping problem is analytically ill-defined, as we need a priori knowledge of the Lagrangian coordinates $\mbi q$ at which the displacement field  and the peculiar velocity field \citep[see][]{Kitaura2016} are evaluated.
Thus, an iterative solution is required to solve this problem, as proposed in pioneering works \citep{Yahil1991,Monaco1999}.
This class of iterative mapping schemes were further developed in \citet{Kitaura2012,Kitaura2012b,kitaura_kigen,Hess2013,birth}. One of the key ingredients to achieve higher accuracy on small scales consists of introducing a Bayesian inference framework in the initial conditions reconstruction which takes into account the likelihood of the dark matter tracers. In this way, the typical additional Gaussian smoothing can be avoided and the number counts of objects within the mesh in a given voxel resolution can be correctly treated \citep{Kitaura2008,Kitaura2010,Jasche2010}. This allows for high precision on a few Mpc scales  \citep[see][]{Kitaura2mrs,Nuza2014}.

\texttt{COSMIC BIRTH} \citep{birth} is a Bayesian inference framework, using  a nested Gibbs-sampling scheme to infer the initial density perturbations $\mbi\delta(\mbi q)$ on a regular cubical mesh grid with $N_{\rm C}$ voxels. It maps the galaxy distribution represented by its Cartesian (for a given set of cosmological parameters) Eulerian redshift-space positions \{$\mbi s^{\rm obs}$\} to  Lagrangian real-space coordinates \{$\mbi q$\} expressed as number counts of galaxies per voxel on a regular mesh: $N_{\rm G}(\mbi q)$. For efficiency the displacement field is computed relying on Augmented Lagrangian perturbation theory \citep{alpt} and so effectively solving Equation \ref{eq:phase} (see Section \ref{sec:alpt}). 

In this way we have a precise description of the action of gravity within $\Lambda$CDM on Mpc scales at redshifts larger than one \citep{Neyrinck2013}. To compute the displacement $\mbi \psi(\mbi q)$ and velocity fields $\mbi v_{\rm r}(\mbi q)$ from the initial density field  $\mbi\delta(\mbi q)$ we apply Hamilton Monte-Carlo (HMC) sampling \citep[see][]{DUANE1987216,Jasche2010, Neal1993,Neal2012} with a bias description ${\mbi B}(\mbi\delta(\mbi q))$ and the galaxy counts on the mesh grid $\mbi N_{\rm G}$.  This is key to perform a forward modelling, as we do not obtain the displacement field from the density field defined in Eulerian space, done by inverse approaches \citep[see e.g.][]{Eisenstein2007}.
We account for selection effects of the galaxy survey data, i.e. the survey geometry, the angular and the radial selection functions, within a response operator $\mathbf{R}$ (see Section \ref{sec:survey3} for details). 
The \texttt{COSMIC BIRTH} code  uses a particularly efficient fourth order leap-frog implementation \citep[see][]{Hernandez-Sanchez2019} to solve the Hamiltonian equations of motion.
We recap the gravity model in Section \ref{sec:alpt} following a description of the probabilistic model in more detail in Section \ref{sec:single} and then expand the calculations in Section \ref{sec:multi} to combine multiple surveys.

\subsection{Gravity Model}
\label{sec:alpt}

In this section we recapitulate the analytical model used within the \texttt{COSMIC BIRTH} code to compute the gravitational evolution of the cosmic density field.

We rely on augmented Lagrangian Perturbation Theory (ALPT) to simulate structure formation \citep[details can be found in][]{alpt}.
In this approximation the displacement field  $\mbi\psi(\mbi q)$, mapping a distribution of dark matter particles at initial Lagrangian positions $\mbi q$ to the final Eulerian positions $\mbi x(z)$ at redshift $z$ ($\mbi x(z)=\mbi q+\mbi\psi(\mbi q)$), is split into a long-range $\mbi\psi_{\rm L}(\mbi q)$ and a short-range component $\mbi\psi_{\rm S}(\mbi q)$, i.e.
$\mbi\psi(\mbi q)=\mbi\psi_{\rm L}(\mbi q)+\mbi\psi_{\rm S}(\mbi q)$.
The long-range component is computed with second order  Lagrangian Perturbation Theory (2LPT) $\mbi\psi_{\rm 2LPT}$
 \citep[for details on 2LPT see][]{BJCP, bouchet1994perturbative,Catelan1995}.
 The resulting displacement field is convoluted with a kernel $\mathcal K$: $\mbi\psi_{\rm L}(\mbi q)={\cal K}(\mbi q,r_{\rm S}) \ast \mbi\psi_{\rm 2LPT}(\mbi q)$, given by a Gaussian filter ${\cal K}(\mbi q,r_{\rm S})=\exp{(-|\mbi q|^2/(2r_{\rm S}^2))}$, with $r_{\rm S}$ being the smoothing radius.
 The short-range component is modelled with the spherical collapse approximation  $\mbi\psi_{\rm SC}(\mbi q)$ \citep[see][]{Bernardeau1994,2006MNRAS.365..939M,Neyrinck2013}.
 The resulting ALPT displacement field from combining the long and the short range components given by:
\be
\label{eq:disp}
\mbi\psi_{\rm ALPT}(\mbi q)={\cal K}(\mbi q,r_{\rm S}) \ast \mbi\psi_{\rm 2LPT}(\mbi q)+\left(1-{\cal K}(\mbi q,r_{\rm S}) \right)\ast \mbi\psi_{\rm SC}(\mbi q)
\ee
is used to move a set of homogenously distributed particles from Lagrangian initial conditions to the Eulerian final ones. We then grid the particles following a clouds-in-cell scheme and phase space mapping \citep{2012MNRAS.427...61A,2013MNRAS.434.1171H} to produce a smooth density field $\mbi\delta(\mbi r)$.
Some improvements can be obtained preventing voids within larger collapsing regions, which essentially extends these regions towards moderate underdensities \citep[see \textsc{muscle} method in][]{2016MNRAS.455L..11N}.
This approach requires about eight additional convolutions being about twice as expensive, as the approach used here.
Moreover, we have checked that the improvement provided by including \textsc{muscle} is not perceptible when using grids with cell resolutions of the order $\sim\mperh$.

The mapping between Eulerian real space $\mbi x(z)$ and redshift space $\mbi s(z)$ is given by: $\mbi s(z)=\mbi r(z)+\mbi v_r(z)$, with $\mbi v_r\equiv(\mbi v\cdot\hat{\mbi r})\hat{\mbi r}/(Ha)$; where  $\hat{\mbi r}$ is the unit sight line vector, $H$ the Hubble constant, $a$ the scale factor, and $\mbi v=\mbi v(\mbi x)$ the 3-d velocity field interpolated at the position of each halo in Eulerian-space $\mbi r$ using the displacement field $\mbi\psi_{\rm ALPT}(\mbi q)$.
We split the peculiar velocity field into a coherent $\mbi v^{\rm coh}$ and a (quasi) virialized component $\mbi v_{\sigma}$: $\mbi v=\mbi v^{\rm coh}+\mbi v^{\sigma}$.
The coherent peculiar velocity field is computed in Lagrangian-space from the linear Gaussian field $\delta^{(1)}(\mbi q)$ using the ALPT formulation consistently with the displacement field (see Equation \ref{eq:disp}):
\be
\mbi v_{\rm ALPT}^{\rm coh}(\mbi q)={\cal K}(\mbi q,r_{\rm S}) \ast \mbi v_{\rm 2LPT}(\mbi q)+\left(1-{\cal K}(\mbi q,r_{\rm S}) \right)\ast \mbi v_{\rm SC}(\mbi q)\,,
\ee
with $\mbi v_{\rm 2LPT}(\mbi q)$ being the second order and $\mbi v_{\rm SC}(\mbi q)$ being the spherical collapse component \citep[for details see][]{Kitaura2014}.
We use the high correlation between the local density field and the velocity dispersion to model the displacement due to (quasi) virialized motions. Effectively, we sample a Gaussian distribution function ($\mathcal G$) with a dispersion \citep[see also][]{Ata2016} given by $\sigma_v\propto\left(1+\delta\left(\mbi r\right)\right)^\gamma$. Consequently we assume,
\be
\mbi v^{\sigma}_r\equiv(\mbi v^{\sigma}\cdot\hat{\mbi r})\hat{\mbi r}/(Ha)\curvearrowleft{\mathcal G}\left(g\times\left(1+\delta(\mbi r)\right)^\gamma\right)\hat{\mbi r}\,.
\ee
For the Gaussian streaming model see \citet[][]{2011MNRAS.417.1913R}, for non-Gaussian models see e.g.~\citet[][]{2007MNRAS.374..477T}. In closely virialised systems the kinetic energy approximately equals the gravitational potential and a Keplerian law predicts $\gamma$ close to $0.5$, leaving only the proportionality constant $g$ as a free parameter in the model. We leave a detailed investigation of the impact of redshift space distortions for future work.

\subsection{Probabilistic Model}
\label{sec:single}

\begin{table*}
\begin{tabular}{|l |l |l |}
\hline
Inferred quantity    & Parent quantity       & Connected via  \\ \hline \hline
\{$\mbi r$\}   \scriptsize{Eulerian real-space}&  \{$\mbi s^{\rm obs}$\} \scriptsize{Eulerian redshift-space}  & $\mbi v_{\rm r}(\mbi q)$ \scriptsize{Peculiar velocity} \\ \hline
\{$\mbi q$\}   \scriptsize{Lagrangian real-space} &  \{$\mbi r$\}  \scriptsize{Eulerian real-space} &  $\mbi \psi (\mbi q)$ \scriptsize{Displacement field}  \\ 
\hline
$\mathbf{R}(\mbi q)$ \scriptsize{Lagrangian response function}  &  $\mathbf{R}(\mbi s)$ \scriptsize{Eulerian response function} &  $\mbi \psi (\mbi q)$, $\mbi v_{\rm r}(\mbi q)$\\ 
\hline
$\mbi \lambda(\mbi q)$ \scriptsize{Galaxy number expectation}& $\mbi N_{{\rm G}}(\mbi q)$ \scriptsize{Galaxy number counts} & $\mbi f_{\bar N}(\mbi q)$, ${\mbi B}(\mbi \delta(\mbi q))$, $\mathbf{R}(\mbi q)$ \scriptsize{(see caption)} \\ 
\hline 
$f_b(z)$  \scriptsize{Bias correction} & $b(z)$  \scriptsize{Linear bias} & $\mbi B(\mbi\delta(\mbi q))$, $\mbi N_{\rm G}(\mbi q)$,$\mathcal{K}(r_{\rm S})$  \scriptsize{(see caption)} \\ 
\hline 
$\mbi\delta(\mbi q)$ \scriptsize{Lagrangian density} &  \{$\mbi q$\} \scriptsize{Lagrangian real-space} &  $\mathcal{P}(\mbi\delta (\mbi q)\mid\mbi N_{\rm G}(\mbi q), {\mbi B}(\mbi\delta(\mbi q)), \mathbf{R}(\mbi q), \mathbf C_{\rm L}(\mbi q) )$ \scriptsize{(see caption)} \\ 
\hline
\end{tabular}
\caption{Inferred quantities of \texttt{COSMIC BIRTH}: The Eulerian real-space positions \{$\mbi r$\}  are inferred from the observed redshift-space positions \{$\mbi s^{\rm obs}$\} via the peculiar velocity $\mbi v_{\rm r}$. The Lagrangian positions \{$\mbi q$\} are calculated from the Eulerian ones by applying the displacement field $\mbi \psi(\mbi q)$. The same mapping is used to calculate the response operator $\mathbf{R}(\mbi q)$ in Lagrangian space. The expectation value of galaxy number counts $\mbi \lambda$ is estimated from the galaxy counts $\mbi N_{{\rm G}}(\mbi q)$, connected by the normalisation factor $\mbi f_{\bar N}(\mbi q)$, the bias model ${\mbi B}(\mbi\delta(\mbi q))$  and the response function $\mathbf{R}(\mbi q))$. We sample the initial density density field $\mbi\delta(\mbi q)$ at  Lagrangian coordinates $\mbi q$ from the posterior probability function $\mathcal{P}(\mbi\delta (\mbi q)\mid\mbi N_{\rm G}(\mbi q),{\mbi B}(\mbi\delta(\mbi q)),\mathbf{R}(\mbi q),\mathbf C_{\rm L}(\mbi q))$. The connection quantities $\mbi v_{\rm r}, \mbi \psi(\mbi q), {\mbi B}(\mbi\delta(\mbi q)), \mathbf{R}(\mbi q), \mathbf C_{\rm L}(\mbi q)$ depend on a set of cosmological parameters $\{p_{\rm c}\}$. $f_b$ is the bias correction term, derived from the large-scale linear bias $b$ and a smoothing kernel $\mathcal{K}$ with radius $r_{\rm S}$. Note that, $\mbi \lambda(\mbi q)$, $\mbi N_{\rm G}(\mbi q)$, ${\mbi B}(\mbi\delta(\mbi q))$, $\mbi f_{\bar N}(\mbi q)$, and $\mbi \delta(\mbi q)$ are  arrays of scalar quantities of $N_{\rm C}$ entries, while $\mbi v_{\rm r}(\mbi q)$, $\mbi \psi(\mbi q)$ are arrays of three-dimensional vector quantities of $N_{\rm C}$ entries. The quantities $\mathbf{R}(\mbi q)$ and $\mathbf C_{\rm L}(\mbi q)$ are matrix operators of $N_{\rm C}\times N_{\rm C}$ dimensionality.}
\label{tab:birth}
\end{table*}

We use a Bayesian framework to draw samples of the density field $\mbi\delta(\mbi q)$ from a posterior probability density $\mathcal{P}(\mbi\delta (\mbi q) \mid \mbi N_{\rm G}(\mbi q),{\mbi B}(\mbi\delta(\mbi q)),\mathbf{R}(\mbi q) )$. The posterior itself is a product of a prior and a likelihood function which we will describe in more detail in the following. 
We express the expectation value of galaxies per voxel $\mbi \lambda_i=\left\langle\mbi N_{{\rm G}}(\mbi q)\right\rangle_i$\footnote{The expectation value is given by the ensemble average: $\langle \mbi X\rangle$.} for all voxels $i\in [1\dots N_{\rm C}]$ as:
\ba
      \lambda(\mbi q)_i = \left\langle\mbi N_{{\rm G}}(\mbi q)\right\rangle_i = f_{{\bar N}i}(\mbi q) \sum_k R_{ik}(\mbi q) { B}_k(\mbi\delta(\mbi q)) \, ,
    \label{eq:expect}
\ea
with the normalisation factor $\mbi f_{\bar N}(\mbi q)={\bar N}/\langle {\mbi B}(\mbi\delta(\mbi q))\rangle$ ensuring the right galaxy number density $\bar N$, as given by each survey. In our case, the response function $\mathbf{R}$ is limited to the completeness $w_i$ in each cell $i$ as described in Section \ref{sec:survey3}: $R_{ij}(\mbi q)= R_{ii} \delta^{\rm K}_{ij}$.
We then relate the expected number counts of galaxies per voxel to the actually observed number counts $\mbi N_{{\rm G}}$ with a Poisson likelihood \citep{Kitaura2008,Kitaura2010} model:
\ba
\label{likelihood}
\mathcal{L} (\mbi N_{{\rm G}}\mid \mbi \lambda(\mbi    q))=\prod_{i}\frac{\lambda_{i}(\mbi q)^{N_{{\rm G}i}}\exp\left({-\lambda_{i}(\mbi q)}\right)}{ N_{{\rm G}i}!}\,,
\ea
where $\mbi \lambda (\mbi q)=\mbi \lambda\left(f_{\bar N}(\mbi q),{\mbi B}(\mbi\delta(\mbi q)),\mathbf{R}(\mbi q)\right)$.
This is an adequate assumption for  tracers  with a vanishing small scale clustering, which otherwise become sources of super-Poissonity \citep[e.g.][]{Peebles1980} that can be modelled with a negative-binomial likelihood \citep[][]{Kitaura2014,Neyrinck2014,Ata2015}. However, since the tracers are mapped to Lagrangian space at very high redshifts (e.g.: $z=100$), a deviation from Poissonity becomes insignificant except for the most massive galaxies \citep{Modi2017,Abidi2018,Schmittfull2019}, which is not the case for our galaxy samples \citep[][]{cosmos2015} (see Section \ref{sec:surveys}).

The above mentioned conditions are ideal to describe the matter distribution with a lognormal prior towards high redshifts \citep{Coles1991}. Its derivation is precisely based on a comoving framework at initial cosmic times, before shell crossing occurs. We apply a logarithmic transformation, which further linearises the density field \citep{Neyrinck2009} as: 
\ba
\mbi\delta_{\rm L}(\mbi q)=\log(1+\mbi\delta(\mbi q))-\mbi\mu\,,
\ea
with $\mbi\mu = \langle \log(1+\mbi\delta)\rangle=-\log\left({\left\langle \rm e^{\mbi \delta_{\rm L}}\right\rangle}\right)$  \citep{2012MNRAS.420...61K}.
Since we consider early cosmic times, the overdensity field has little power $|\delta|\ll1$. Hence, the logarithmic transformation ensures positive densities $\mbi \rho$ (with $\mbi \delta=\mbi \rho/\bar{\rho}-1=\exp(\mbi \delta_{\rm L}+\mbi \mu)-1$), while we can model the prior $\pi(\mbi\delta(\mbi q))$\footnote{Note that the prior is actually a function of the linearised density field $\mbi\delta_{\rm L}(\mbi q)$, which is in turn a function of the original density field $\mbi\delta(\mbi q)$.} for the linear density field $\mbi \delta_{\rm L}$ by a Gaussian distribution with zero mean
\begin{eqnarray}
\label{eq:prior}
\lefteqn{\pi(\mbi\delta(\mbi q)\mid \mathbf C_{\rm L}(\mbi q))=}
\nonumber\\&&\frac{1}{\sqrt{(2\pi)^{N_{\rm C}}\det(\mathbf C_{\rm L}(\mbi q))}}\exp\left(-\frac{1}{2}\mbi \delta^{\dagger}_{\rm L}(\mbi q)\mathbf C_{\rm L}^{-1}(\mbi q)\mbi\delta_{\rm L}(\mbi q)\right)\,,
\end{eqnarray}
where $\mathbf C_{\rm L}(\mbi q)=\left\langle\mbi\delta^{\dagger}_{\rm L}(\mbi q)\mbi\delta_{\rm L}(\mbi q)\right\rangle$ is the covariance matrix of the linearised density fields, which depends on the cosmological parameters $\{p_{\rm c}\}$.
Finally, we can express the posterior of $\mbi\delta(\mbi q)$ through Bayes theorem as:
\ba
\lefteqn{\mathcal{P}(\mbi\delta (\mbi q)\mid\mbi N_{\rm G}(\mbi q),{\mbi B}(\mbi\delta(\mbi q)),\mathbf{R}(\mbi q),\{p_{\rm c}\} ) \propto}\nonumber \\ &\pi\left(\mbi\delta(\mbi q)\mid \mathbf C_{\rm L}(\{p_{\rm c}\})\right) \times \mathcal{L}(\mbi N_{\rm G}((\mbi q))\mid\mbi\lambda(\mbi q),\mathbf{R}(\mbi q) )\, ,
\label{eq:post}
\ea
where the normalisation given by the evidence is not necessary within HMC.
Table \ref{tab:birth} summarises the main quantities that are sampled and how they are connected to each other.

\subsubsection{Galaxy Bias Description}
\label{sec:bias}

Finally, we need to specify the connection between the likelihood and the prior through the bias relation ${\mbi B}(\mbi\delta)$.
In the \texttt{COSMIC BIRTH}  framework, non-local bias is described through the displacement. The split-background bias \citep{Kaiser1984,bbks1986}, which is necessary in Eulerian space particularly for massive galaxies \citep[see e.g.][]{Kitaura2015}, however, becomes negligible when homogenising the galaxy distribution mapping it to Lagrangian space. 
Hence, we can assume a power-law Lagrangian bias as discussed in \citet[][]{birth}:
\ba
\label{eq:lagbias}
{\mbi B}(\mbi\delta(\mbi q))=(1+\mbi\delta(\mbi q))^{b(z_q)\,f_b(z_q)}\,,
\ea
where $z_q$ is the redshift at which the Lagrangian coordinates are evaluated (for this study $z_q=100$), $b$ the linear large-scale bias and $f_b$ the non-linear correction factor of our bias description \citep{Ata2016}. The correction factor $f_b(z_q)$ can be determined iteratively as presented in \citet{birth}, ensuring that $b$ exactly corresponds the large-scale linear bias. 
By using a power-law bias we ensure that the density field is positive, since otherwise any bias less than one can potentially cause negative densities at voxels with $\delta$ close to $-1$.

The advantage of this bias description is that the only free parameter of our method is reduced to the large-scale bias at Eulerian space, which can be connected to Lagrangian space through passive evolution \citep[][]{Nusser1994,Fry1996}:
\ba
\label{eq:passive}
b(z_q) = (b(z)-1) \frac{D(z)}{D(z_q)}+1 \, ,
\ea
including the linear growth factor $D(z)$ .
We will show in Section \ref{sec:lsbiasc} how we describe the large-scale bias evolution for the employed galaxy catalogs.

\subsection{Summary of the Algorithm}

In summary, the joint probability distributions of all the above mentioned variables can be expressed within a Gibbs-sampling scheme based on the corresponding conditional probabilities:
\ba
\label{eq:inference}
\delta(\mbi q)&\curvearrowleft& \mathcal{P}_\delta(\mbi\delta (\mbi q)\mid\mbi N_{\rm G}(\mbi q), {\mbi B}(\mbi\delta(\mbi q)), \mathbf{R}(\mbi q), \mathbf C_{\rm L}(\mbi q) ) \nonumber \\
\{\mbi r\}&\curvearrowleft& \mathcal P_{r}\left(\{\mbi r\}|\{\mbi s^{\rm obs}\},\mbi v_{\rm r}(\mbi q),{\cal M}_{v}\right)\,,\nonumber\\
{\{\mbi q\}}&\curvearrowleft& \mathcal P_{q}\left(\{\mbi q\}|\{\mbi r\},\mbi \psi({\mbi q}),{\cal M}_{\psi}\right)\,,\nonumber\\
\mathbf{R}(\mbi q)&\curvearrowleft& \mathcal P_{R}\left(\mathbf{R}(\mbi q)|\mathbf{R}(\mbi s),\mbi \psi({\mbi q}),{\cal M}_{\psi}\right)\,,\nonumber\\
{\mbi B}(\mbi\delta(\mbi q))&\curvearrowleft& \mathcal P_{B}\left({\mbi B}(\mbi\delta(\mbi q))|{\mbi B}(\mbi\delta(\mbi s)),\mbi \psi(\mbi q),{\cal M}_{\psi}\right)
\,{,}
\ea
where the functional dependency of $\mbi q$ and $\mbi s$ stand for Lagrangian real-space, and Eulerian redshift-space coordinates, respectively. The curved left arrows stand for the sampling process. ${\cal M}_{v}$ and ${\cal M}_{\psi}$ represent the models describing peculiar motions and displacement fields.

\subsection{Multi-tracer Formalism in Lagrangian-space}
\label{sec:multi}
In this study, we aim at combining the data of five spectroscopic surveys in the COSMOS field that share spatially overlapping footprints and similar redshift distributions, however relying on different observing strategies (for more details see Section \ref{sec:surveys}).
Thus merging the surveys into a single catalogue by a pre-processing step, i.e. adding the different catalogues into one data set, represents a difficult task, which in general cannot be accomplished without making a series of simplifying assumptions, e.g. neglecting the different selection criteria.

Some of the previous pioneering Bayesian inference studies have already applied multi-tracer treatment, however, all performed in Eulerian space, and without taking into account separate survey footprints  within the Bayesian  scheme \citep{Jascheborg, Granettvimos}, or separate footprints from the same (super)-set of catalogs \citep{Jasche2017, Jasche2019}. 

In this work, we aim at performing joint analysis of entirely different surveys within the Bayesian framework.  According to Section  \ref{sec:single}, we perform the reconstructions in Lagrangian space (in our case corresponding to a redshift of $z=100$), where gravitational interactions have not yet introduced mode couplings of density perturbations. Therefore, mapping the galaxies to Lagrangian space before the density sampling step (see Table  \ref{tab:birth}) is gradually reducing the covariance of the different surveys and homogenizes the galaxy fields. This enables us to treat each survey as a distinct component, avoiding mixed terms in the likelihood expression. Therefore, spatial overlap in Lagrangian space does not represent a problem, as long as we make sure that no galaxy is multiply counted among the different surveys.

In this way, we are able to combine different catalogues (indexed with superscript $k$) taking into account  their distinct survey selection functions $\mathbf{R}^k(\mbi q)$, number densities $\mbi N^k_{\rm G}(\mbi q)$,  and galaxy bias functions ${\mbi B}^k(\mbi\delta(\mbi q))$. 


The Eulerian to Lagrangian mapping of \texttt{COSMIC BIRTH}, shown in Equation \ref{eq:phase}, can lead to a change of the redshift bin of a tracer. Consequently, the bias of this tracer will not coincide with the bias of the redshift bin at its new location. This can be taken into account by keeping track of galaxies staying at a redshift bin, or jumping from one redshift bin to another,  which causes a ``bias mixing'' implemented in the \texttt{COSMIC BIRTH}  code  \citep[see Section 3 in][]{birth}. While this effect is negligible when  interpolating the bias within the redshift bins, it has the  advantage that a  multi-tracer treatment is already implemented in this framework. 
We can then extend  \texttt{COSMIC BIRTH} to perform a full Bayesian multi-tracer \& multi-survey analysis to address the challenges of this work following the calculations presented in e.g. \citet[Appendix A]{Ata2015} to express the corresponding posterior $\mathcal{P}^{\rm multi}$ in Lagrangian coordinates and combine the surveys with their specific likelihoods to construct the combined likelihood $\mathcal{L}^{\rm multi}$ by:
\ba
\lefteqn{\mathcal{L}^{\rm multi}\left(\mbi N_{\rm G}(\mbi q)|\mbi \lambda\left(f_{\bar N}(\mbi q),{\mbi B}(\mbi\delta(\mbi q)),\mathbf{R}(\mbi q)\right) \right) \propto}\\&& \prod_k \mathcal{L}^{(k)}\left(\mbi N^{(k)}_{\rm G}(\mbi q)|\mbi \lambda^{(k)}\left(f_{\bar N}^{(k)}(\mbi q),{\mbi B}^{(k)}(\mbi\delta(\mbi q)),\mathbf{R}^{(k)}(\mbi q)\right)\right)\nonumber\, ,
\ea
where index $k\in[1\dots N_{\rm S}]$ denotes the different surveys.
Accordingly, we need to generalise Equation \ref{eq:expect} and distinguish between the surveys in the reconstructed volume:
\ba
   \label{eq:expect_multi}
     \mbi \lambda^{(k)} = \left\langle \mbi N^{(k)}_{\rm G} (\mbi q)\right\rangle = f^{(k)}_{\bar N}(\mbi q) \mathbf{R}^{(k)}(\mbi q) {\mbi B}^{(k)}( \mbi\delta(\mbi q)) \, .
\ea
In Hamiltonian sampling, one seeks to draw samples of the potential energy term $\mathcal{U}$ of the Hamiltonian, that is linked to the posterior in Equation \ref{eq:post} via 
\ba
\mathcal{U}=-\ln{\mathcal{P}} \, . 
\label{eq:potential}
\ea
Thus, the multi-tracer \& multi-survey posterior $\mathcal{P}^{\rm multi}$ writes for a number of $N_{\rm S}$ surveys as:
\ba
\label{eq:multi_post}
\lefteqn{-\ln{\mathcal{P}^{\rm multi}}\left(\mbi\delta(\mbi q) |\mbi \lambda^{(1)}(\mbi q), \mbi \lambda^{(2)}(\mbi q), \dots, \mbi \lambda^{(N_{\rm s})}(\mbi q)\right) =} \\ && c-\ln{\pi(\mbi\delta(\mbi q)\mid \mathbf C_{\rm L}(\mbi q))} \nonumber \\&& -\ln\mathcal{L}^{(1)}\left(\mbi N^{(1)}_{\rm G}|\mbi \lambda^{(1)}\left(f_{\bar N}^{(1)}(\mbi q),{\mbi B}^{(1)}(\mbi\delta(\mbi q)),\mathbf{R}^{(1)}(\mbi q)\right)\right)\nonumber \\&& -\ln\mathcal{L}^{(2)}\left(\mbi N^{(2)}_{\rm G}|\mbi \lambda^{(2)}\left(f_{\bar N}^{(2)}(\mbi q),{\mbi B}^{(2)}(\mbi\delta(\mbi q)),\mathbf{R}^{(2)}(\mbi q)\right)\right)\nonumber\\ && \vdots \nonumber\\ && -\ln\mathcal{L}^{(N_{\rm S})}\left(\mbi N^{(N_{\rm S})}_{\rm G}|\mbi \lambda^{(N_{\rm S})}\left(f_{\bar N}^{(N_{\rm S})}(\mbi q),{\mbi B}^{(N_{\rm S})}(\mbi\delta(\mbi q)),\mathbf{R}^{(N_{\rm S})}(\mbi q)\right)\right) \nonumber\, ,
\ea
where the constant $c$ does not depend on $\mbi\delta(\mbi q)$.

\section{Survey Data \& Completeness}
\label{sec:surveys}
Within the COSMOS field, several spectroscopic surveys have been undertaken \citep[see e.g.][for a summary]{2018ApJ...858...77H}, focusing mainly on star forming galaxies at high redshifts. This work uses data from five different surveys, summarised in Table \ref{tab:surveys}.  
\begin{table}
\begin{tabular}{ |l|c|c|l| }
\hline 
Survey  & $N^{\rm Obj}$      & $z$ range                & Parent catalog   \\ \hline \hline
zCOSMOS-deep &   3544        &  $1.4 \leq z \leq 3.6$(*)  & COSMOS      \\ \hline
VUDS         &   1822        &  $1.4 \leq z \leq 3.6$(*)  & COSMOS      \\ \hline
MOSDEF       &   401         &  $1.4 \leq z \leq 3.6$(*)  & 3D-HST      \\ \hline
ZFIRE        &   149         &  $2.0 \leq z \leq 2.2$   & ZFOURGE     \\ \hline
FMOS-COSMOS  &   587         &  $1.4 \leq z \leq 1.7$   & COSMOS      \\ \hline 
\end{tabular}
\caption{Summary of five surveys used for this study. $N^{\rm Obj}$ resembles the number of galaxies that we use after applying spectroscopic quality criteria and removing duplicates. (*) zCOSMOS-deep, VUDS \& MOSDEF exceed the redshift range of our reconstructions. MOSDEF observes the redshift range in intervals of $1.37 \leq z \leq 1.70$ , $2.09 \leq z \leq 2.61$, and $2.95 \leq z \leq 3.80$. In the case of FMOS-COSMOS, about half of the galaxies in the central footprint are used in this work.}
\label{tab:surveys}
\end{table}
We describe each survey in more detail below and afterwards explain our method to estimate the corresponding survey completenesses.

\subsection{Surveys in the COSMOS Field}
\label{sec:survey2}
Let us briefly recap the main characteristics of the different spectroscopic galaxy surveys considered in this work.

\begin{figure}
    \centering
    \hspace*{-.3cm}
    \includegraphics[width=0.49\textwidth]{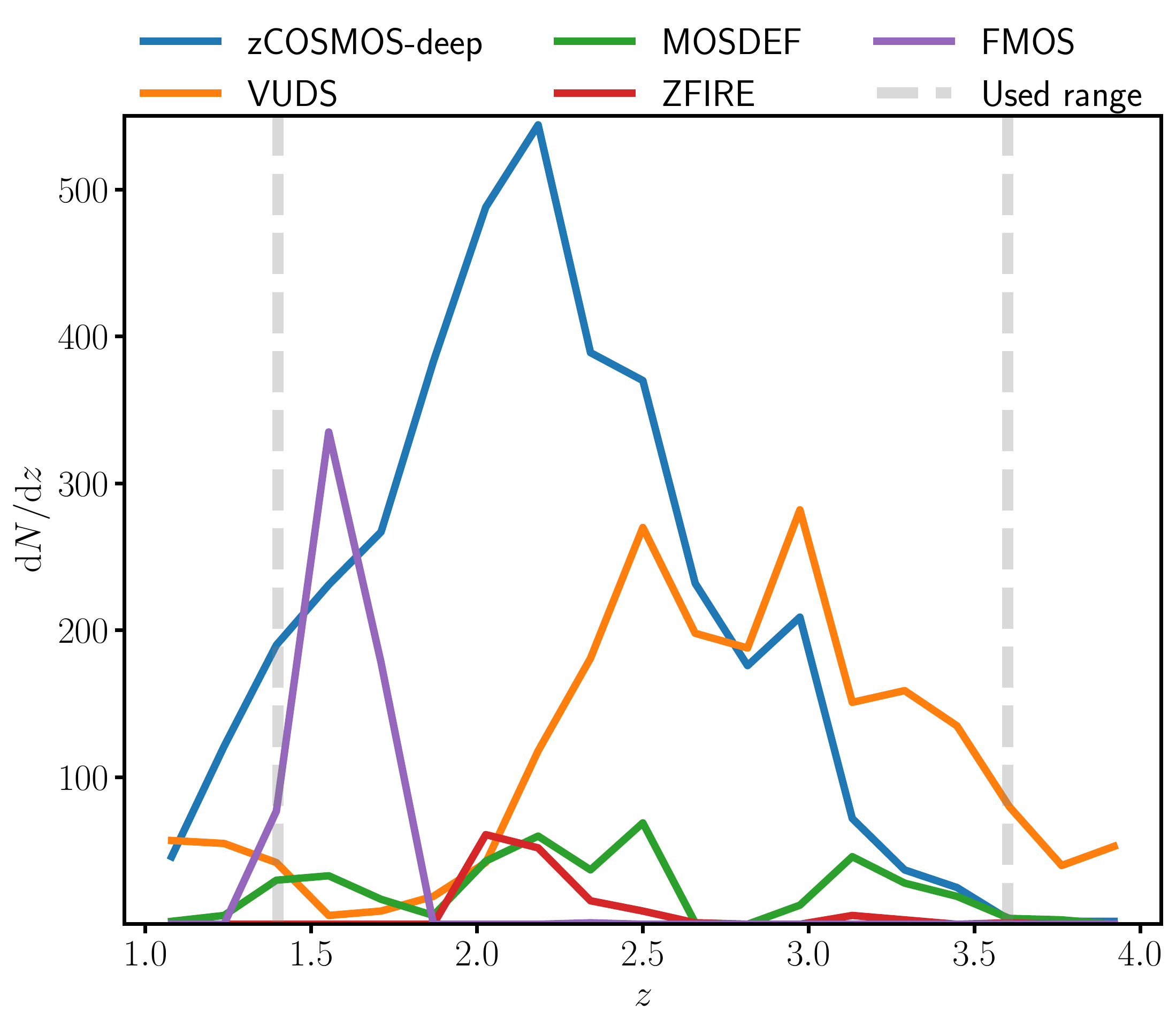}
    \caption{Radial distribution functions for the five considered surveys shown as function of redshift with a bin width of $\Delta z = 0.16$. The two grey dashed vertical lines signpost the considered redshift range $1.4 \leq z \leq 3.6$ in this work.}
\label{fig:nz}
\end{figure}

\begin{itemize}
    \item \textbf{zCOSMOS-deep}\\ 
    The zCOSMOS-deep survey is the high-redshift component of the zCOSMOS spectroscopic survey, which covers the central $1~\mathrm{deg}^2$ of the zCOSMOS footprint \citep[][Lilly et al in prep.]{Lilly2006} using the VIMOS spectrograph \citep{LeFevre2003} on the VLT. The targets were chosen from the then-current version of the multi-colour photometric COSMOS catalog \citep{Capak2007}. To isolate galaxies at redshifts of $z>1.5$, several selection criteria were applied. In particular,  a  selection in the (U-B)/(V-R) colour-colour plane, called ``UBR'' selection \citep{Steidel2004} was combined with the ``BzK'' selection \citep{Daddi_2004}. For both the ``UGR'' and ``BzK'' selections, an additional selection of $22.5 < B_{\rm{AB}} < 25.0$ and a deep $K$-band imaging reaching down to $K_{\rm{AB}}\sim 23.5$ were applied. Our analysis is done on a tentative version of the zCOSMOS-deep catalog that has been used for the bulk of previous works that employ zCOSMOS-deep data. A refined version of this catalog will be available in the future (Lilly et al. in prep).  \\
    \item \textbf{VUDS} \\ 
    The VIMOS Ultra Deep Survey \citep{LeFevre2015}, hereafter VUDS, is another spectroscopic survey carried out on the VIMOS spectrograph that was partly operated in the COSMOS field, but considerably deeper than zCOSMOS-deep with up to $\sim 3\times$ larger integration times. VUDS peaks in number density at $z\sim 3$ and was designed to explore multiple questions, including those related to the formation rates of stars and merging of galaxies during the period of time when galaxies were most active. Another major success was to identify and characterize galaxy protoclusters at $z\sim 2-4$ \citep[e.g.][]{Cucciati2014, Lemaux2018, Cucciati2018}. The selection is based on photometric redshift selections \citep{Ilbert2013}, with a small fraction of galaxies selected using the Lyman-break technique \citep{Steidel1996}. VUDS and zCOSMOS are based on different versions of COSMOS parent catalog, so the astrometry was not identical for all sources between the two surveys. Because of the different catalogs used, we employed a matching radius of $0.1-0.2\arcsec$ to identify true duplicates.\\

    \item \textbf{MOSDEF} \\ 
    The MOSFIRE Deep Evolution Field (MOSDEF) Survey \citep{Kriek2015} is partly taken in the COSMOS field, separated into three redshift intervals at $1.37 \leq z \leq 1.70$, $2.09 \leq z \leq 2.61$, and $2.95 \leq z \leq 3.80$, down to fixed $H_{\mathrm{AB}}$ magnitudes of $24.0$, $24.5$, and $25.0$ for each interval. The MOSFIRE spectrosgraph \citep{10.1117/12.924794} on the Keck-I telescope was used to obtain near infra-red emission line redshifts of the targeted galaxies. The targets were selected from the photometric and grism 3D-HST \citep{Brammer2012} data, applying photometric redshift and magnitude requirements to the parent catalog and show a spectroscopic success rate of around $\sim80\%$. \\
    
    \item \textbf{ZFIRE} \\
    The KECK/MOSFIRE Spectroscopic Survey of Galaxies in Rich Environments at $z\sim 2$ (ZFIRE) \citep{ZFIRE2016} using the MOSFIRE spectrosgraph was partly taken in the COSMOS and Hubble Ultra Deep Survey (UDS) field \citep{Beckwith_2006}. For the COSMOS field the targets were $K$-band selected from the photometric parent FourStar Galaxy Evolution Survey (ZFOURGE) catalog \citep{ZFOURGE2016}, requiring a photometric redshift $2.0 \lesssim z_{\rm phot} \lesssim 2.2$. ZFIRE was designed to observe primarily the $z = 2.095$ galaxy cluster \citep{cluster1}.  \\
    
    \item \textbf{FMOS-COSMOS} \\
    The FMOS-COSMOS survey \citep[e.g.][]{Silverman2015,FMOS} (hereafter FMOS) used the Fibre Multi-Object Spectrograph \citep{FMOS-spec} at the Subaru Telescope, observing star forming galaxies at redshifts $z\sim 1.6$ in the near infra-red. The targets were selected from the COSMOS photometric catalog within a redshift range of $1.4 \lesssim  z_{\rm phot} \lesssim 1.7$, additionally applying limits on the UltraVISTA K-band magnitude limit and the $H\alpha$ flux predicted from SED fitting. FMOS objects are pre-selected with secure photometric redshift using the filters available in the COSMOS photometric survey. In this analysis we utilize FMOS observations in the range of $149.8 \leq \mathrm{R.A.} \leq 150.4$ and  $1.8 \leq \mathrm{DEC}\leq2.5$.

\end{itemize}
In summary, the surveys apply different colour, magnitude and photometric redshift pre-selection cuts to the parent photometric catalogs to efficiently select high redshift targets for spectroscopy. In the case of galaxies that were spectroscopically observed in more than one of these surveys, we kept the one with better redshift quality, based on the signal-to-noise ratio of the observed spectra (see Section \ref{sec:angular}). 
From the redshift distributions of the surveys shown in Figure \ref{fig:nz} we can  see that the number density of galaxies peaks at $2 \leq z \leq 2.5$, mainly contributed by the zCOSMOS-deep survey.

\begin{figure}
    \centering 
    \hspace*{-0.3cm}
    \includegraphics[width=0.49\textwidth]{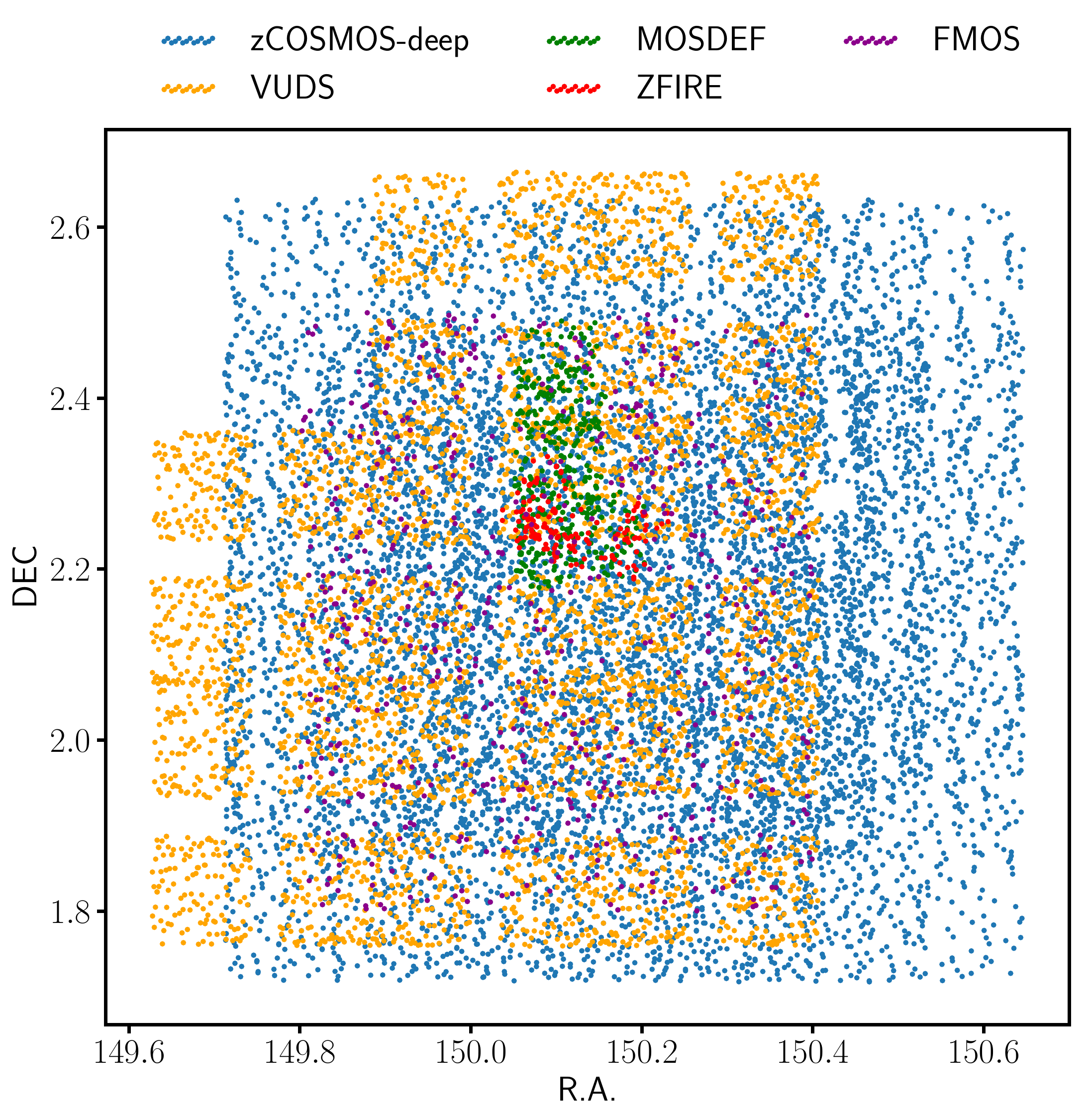}
    \caption{Survey footprints: Spatial distribution of the observed galaxies in the  right ascension ($\mathrm{R.A.}$)-declination ($\mathrm{DEC}$) plane. We follow the same colour coding as in Figure \ref{fig:nz}. At a redshift of $z=2.5$, the maximum separation  $\Delta \mathrm{DEC}=0.91^\circ$ corresponds to a comoving transverse distance of $d_{\Delta\mathrm{DEC}} = 64.8 \mperh$, while for $\Delta \mathrm{R.A.}=1.02^\circ$ the maximum separation is $d_{\Delta\mathrm{R.A.}} = 70.2 \mperh$ assuming the cosmological parameters in Section \ref{sec:intro}.}
\label{fig:scatter_comb}
\end{figure}

The footprints (galaxy distribution in angular coordinates) of each survey are shown in Figure \ref{fig:scatter_comb}.
From this figure we can recognise that the edges of zCOSMOS-deep, represented in blue,  have been observed with only a single pointing, thus having a lower number density, while in the central area the pointings are overlapping, allowing for a denser targeting.
Regarding the VUDS survey, represented in orange, Figure \ref{fig:scatter_comb} shows the gaps in between the quadrants of an individual VIMOS pointing, whereas within one quadrant the target sampling is very homogeneous. Focusing on the MOSDEF and ZFIRE surveys depicted in green and red, respectively, one can see that they cover smaller areas in the central part of the COSMOS field. The FMOS survey, represented in magenta, shows a sparser targeting compared to the rest of the surveys considered in this study. 
While the three surveys MOSDEF, ZFIRE, and FMOS lie withing the footprints of zCOSMOS-deep and VUDS, the latter ones only partly overlap. VUDS shares 86\% angular coverage with zCOSMOS-deep, which corresponds to 69\% VUDS coverage of the zCOSMOS-deep footprint.

\subsection{Survey Completeness Estimation}
\label{sec:survey3}
\begin{figure}
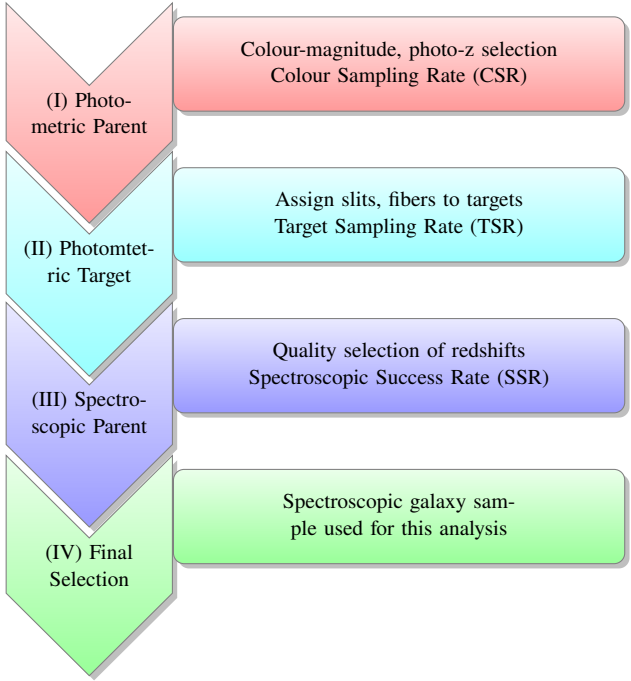

\centering
\tikzset{description title/.append style={
    signal, 
    signal to=south, 
    signal from=north,
    yshift=-0.65cm,
  }
}
\begin{center}
\smartdiagramset{description title width=2.cm, 
description title text width=2cm,
descriptive items y sep=2,
description text width=5.75cm,
module minimum height=1.25cm}
\smartdiagram[descriptive diagram]{%
{(I) Photometric Parent, {Colour-magnitude, photo-z selection \\ Colour Sampling Rate (CSR)}},
{(II) Photomtetric Target, {Assign slits, fibers to targets \\ Target Sampling Rate (TSR)}},
{(III) Spectroscopic Parent, {Quality selection of redshifts \\ Spectroscopic Success Rate (SSR)}},
{(IV) Final Selection, Spectroscopic galaxy sample used for this analysis},}
\end{center}
\caption{Flowchart describing the process to obtain the final spectroscopic catalog, starting from the parent photometric survey. The arrow symbol on the left represents the name of the survey. The descriptive text on the right denotes the selection/operation that is undertaken upon the survey, leading to the next stage. We use an adapted terminology of the VVDS/VIMOS collaboration: Stage (I) to (II) is called  Colour Sampling Rate (CSR), (II) to (III) Target Sampling Rate (TSR), and finally (III) to (IV) Spectroscopic Success Rate (SSR).}
\label{fig:flowchart} 
\end{figure}
The response function $\mathbf{R}$ (see Table \ref{tab:birth}, Section \ref{sec:birth}) represents the efficiency how each voxel $i$ has been observed as compared to the number of possible targets (see Figure \ref{fig:flowchart} for details of the selection stages). In a Bayesian analysis, the likelihood function should therefore account for the uncertainty of the galaxy number counts $\mbi N_{\rm G}(\mbi q)$ as a function of the completeness, shown in Equation \ref{eq:expect}. The response function is calculated from an angular and a radial component, $\mathbf{R}_\alpha$ and $\mathbf{R}_r$, respectively, as they can be independently calculated:
\ba
\label{eq:response}
\mathbf{R} = \mathbf{R}_\alpha \cdot \mathbf{R}_r\, .
\ea
Both components are calculated on the reconstructed mesh grid and then multiplied for each voxel.
It is practical to consider them separately, since the angular part is not subject to redshift space distortions, contrary to the radial part.
Following \citet{birth}, we can compute the angular response operator in Eulerian space $\mathbf{R}_\alpha(\mbi s)=\mathbf{R}_\alpha(\mbi r)$ once, and in each Gibbs-sampling iteration map it to Lagrangian space $\mathbf{R}_\alpha(\mbi q)$ through the displacement field (see Table \ref{tab:birth}).
The radial response function can be trivially computed from the distribution of large-scale tracers in Lagrangian real space coordinates $\{\mbi q\}$ in each Gibbs-sampling iteration, and multiplied according to Equation \ref{eq:response}.
Let us thus focus in detail on the computation of the angular completeness.

\subsubsection{Angular completeness}
\label{sec:angular}
In the following we adopt the terminology of the VIMOS related surveys (e.g. VVDS,VIPERS) \citep[see e.g.][]{2005A&A...439..863I,2009A&A...508.1217Z, Vipers, Granettvimos,Vipers2} to estimate the spatial completeness in the reconstructed volume. 

To compute the angular completeness, we need to distinguish between four selection stages, starting from the photometric parent catalog and yielding the final spectroscopic survey, shown in Figure \ref{fig:flowchart}. 
In the first stage, observers apply photometric selection criteria on top of the \textit{parent photometric catalog} (I) to select galaxies with certain properties (e.g. star forming, redshift range etc). The resulting catalog consists of \textit{photometric targets} (II), leading to a second stage, as shown in Figure \ref{fig:flowchart}. The transition from the photometric parent (I) to the photometric target catalog (II) is called \textit{Colour Sampling Rate} (CSR) and accounts for the colour-colour, colour-magnitude and photometric redshift selections. A fraction of the photometric targets (II) is chosen for spectroscopy, which we call \textit{Spectroscopic Parent} (III). The ratio of (II) and (III) is called \textit{Target Sampling Rate} (TSR). Many factors have an influence on the final selection of spectroscopic galaxies such as the spatial arrangement of the slits/fibres, the conditions during the observations etc. This means, that not all the spectra taken for individual galaxies can be translated into a reliable redshifts. We call the ratio of the spectroscopic parents (III) to the final selection (IV) the \textit{Spectroscopic Success Rate} (SSR). 

The selection estimation approach consists on reproducing the targeting strategy of each survey to precisely estimate the ratio from the parent photometric catalog to the final  spectroscopic galaxies' selection. 
This has been neglected in previous studies based on the COSMOS field (see Section \S \ref{sec:reconstructions} and references therein).

Let us describe the different selection steps in more detail: 
\begin{itemize}
    \item \textbf{Colour Sampling Rate (CSR)}\\ One can make the robust assumption that the photometric pre-selections applied on the parent photometric catalog (I) are constant over the footprint of each survey. In such a case, the CSR will only result in an overall normalization factor of the number density of galaxies \citep[see][]{Pezzotta2017}, but not influence the angular dependent clustering. Therefore, we can safely absorb this factor into the radial selection factor.  
    \item \textbf{Target Sampling Rate (TSR)}\\ First, we reproduce the photometric pre-selection criteria.
    Then we build the ratio of the photometric targets $N_\mathrm{II}$ with the number of the spectroscopic parents $N_\mathrm{III}$, regardless of the quality of the spectra, $\mathrm{TSR} =N_\mathrm{III} / N_\mathrm{II}$. We construct a mesh grid $80\times80$ cells over the R.A.-DEC plane (Figure \ref{fig:scatter_comb}), resulting in a resolution of about $0.9 ~\mathrm{arcmin}$ per cell. We note that the reconstructions are computed with a comoving resolution of $d_\mathrm{R} = 2\mperh$ (see Section \ref{sec:reconstructions}). This corresponds to a angular aperture of $\vartheta = 1.5\arcmin$ at a redshift of $z=3.6$ and thus coarser than the angular completeness resolution. 
    We set the value of the selection function in between the VUDS quadrants and outside the borders of the surveys to zero.
    \item \textbf{Spectroscopic Sampling Rate (SSR)}\\  Finally, we select a subset $N_{\rm IV}$ of galaxies from the spectroscopic parent sample, that have high redshift accuracies, as described follows. 
    For the zCOSMOS-deep and VUDS surveys we demand redshift flags of $\ge 2$.
    For the MOSDEF survey we apply quality flags of $=3$, which corresponds to a redshift confidence of $>95\%$ and a minimum signal-to-noise ratio of $2 \leq S/N \leq 3$. For the ZFIRE survey we apply quality flag of 2, which corresponds to a $S/N \geq 5$ and $|z_{\rm spec} - z_{\rm phot} | \le 0.2$. For the FMOS survey we demand  $3 \leq S/N \le 5$, which translates into a redshift quality flag of $\ge2$. 
    The SSR is then calculated as $\mathrm{SSR} =N_{\rm IV} / N_{\rm III}$. 
\end{itemize}

\begin{figure}
    \centering
    \hspace*{-0.2cm}
    \includegraphics[width=0.51\textwidth]{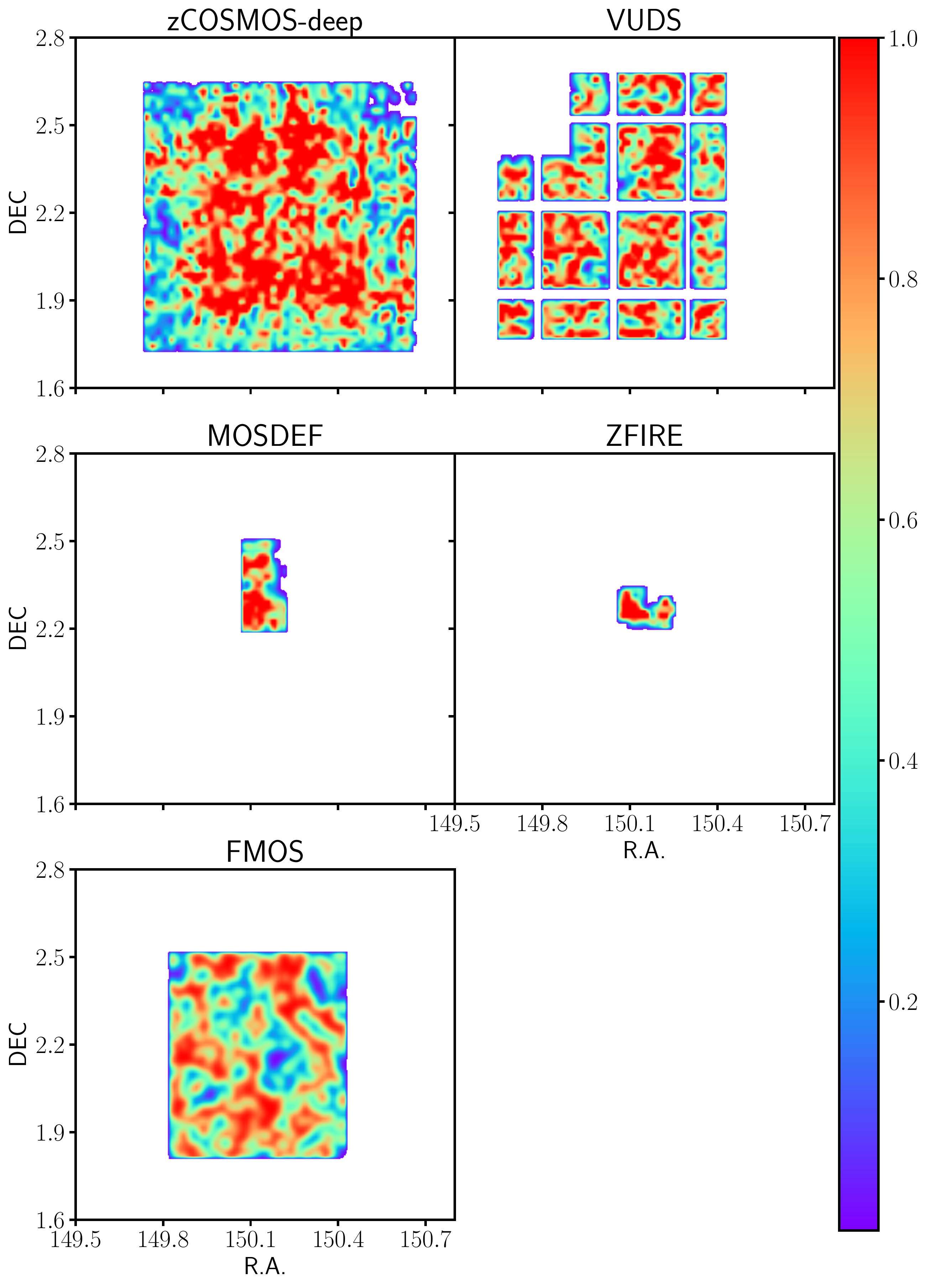}
    \caption{Final angular completeness mask $w_\alpha^k$ for all five surveys computed from TSR and SSR in the R.A.-DEC plane normalized to unity.}
\label{fig:comp_comb}
\end{figure}

\begin{figure*}
    \centering
    \hspace*{-1cm} \vspace*{-.77cm}
    \includegraphics[width=1.1\textwidth]{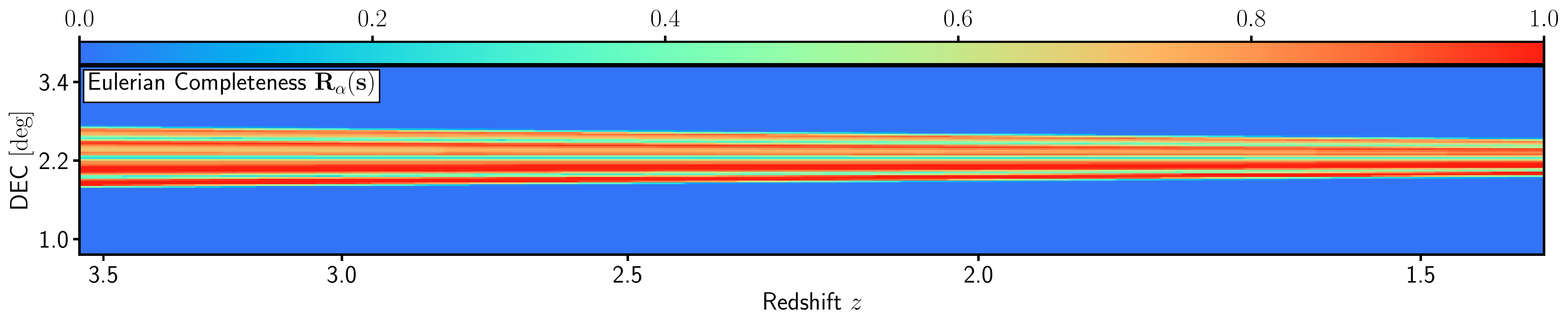}
    \hspace*{-1cm} \vspace*{-.5cm}
    \includegraphics[width=1.1\textwidth]{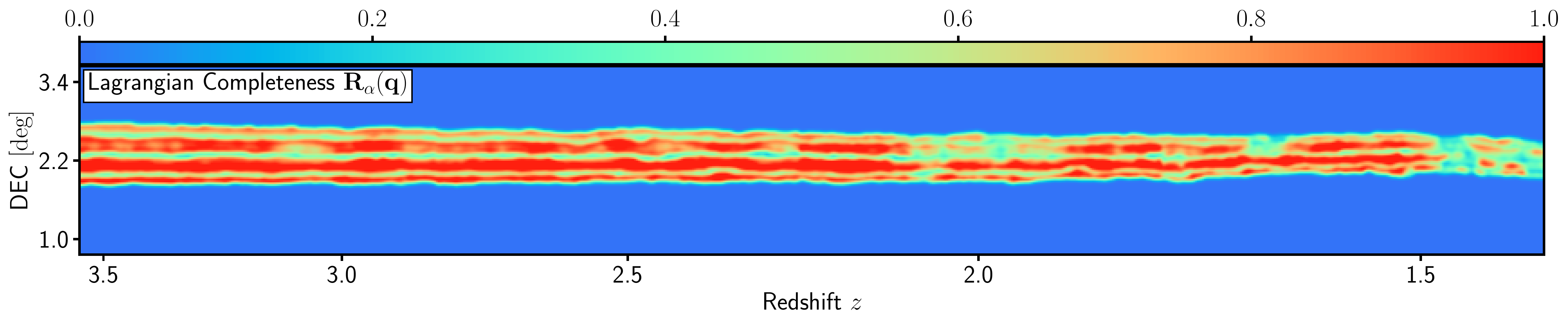}
    \caption{Single slice plot of the projected window function $\mathbf{R}_\alpha$ for the VUDS survey into the reconstructed volume, normalized to unity, showing 0 for non observed regions and 1 for maximum completeness. {\bf The top panel} represents  the window calculated directly from the observations in Eulerian space $\mbi s$, showing the effects of the gaps in between the VUDS quadrants (compare Figure \ref{fig:comp_comb}), emphasizing the importance of an accurate selection function handling for this work. {\bf The bottom panel} presents the same slice in Lagrangian space $\mbi q$. The whole declination angle on the $Y$-axis corresponds to $200 \mperh$ in comoving distance.}
\label{fig:winvuds}
\end{figure*}

The resulting angular selection masks 
\ba
w^k_\alpha = {\rm TSR}^{k} \times {\rm SSR}^{k}\,,
\label{eq:ang_mask}
\ea
in the R.A-DEC plane are shown in Figure \ref{fig:comp_comb} for each survey $k$.
In the case of the zCOSMOS-deep survey the higher targeting rate in the center of the survey footprint, caused by several overlapping pointings with the VIMOS spectrograph, can be appreciated. For the VUDS survey on the contrary, each pointing is unique, and hence,  the areas do not overlap. This results in thin stripes between the quadrants of the pointings in the footprint, showing the inter-CCD gaps of $\sim 2$ arcmin in the VIMOS focal plane. 

Once we have computed the angular mask, we need to project it into three dimensions, as required by the \texttt{COSMIC BIRTH} code.

In particular, we project each R.A.-DEC bin value of $w_\alpha^k$ into our cubical mesh grid on which we perform the reconstructions. As the resolution of $w_\alpha^k$ is higher compared to the reconstruction  mesh grid, we average over each sight at a voxel $i$ and thus obtain the angular completeness $R^k_{ii}$ for all $k$ surveys and cell $i$.
Figure \ref{fig:winvuds} exemplary shows the projection of the VUDS angular selection mask into the reconstructed volume. The data preparation and details of the coordinate system are explained in Section \ref{sec:prep}. The top panel of Figure \ref{fig:winvuds} shows the stripes and the corresponding gaps in between the pointings of the survey in the Eulerian frame $\mathbf{R}_\alpha(\mbi s)$, in accordance with Figures \ref{fig:scatter_comb} and \ref{fig:comp_comb}. The VUDS inter-CCD gaps translate into empty regions of $2.3 \mperh$ at redshifts of $z=3.6$.
The translation of the angular response operator to Lagrangian space $\mathbf{R}_\alpha(\mbi q)$ through the action of gravity causes a deformation in the survey window, which is represented in the lower panel of Figure \ref{fig:winvuds}. This means, that unobserved angular regions in Eulerian space, might have been effectively partially observed in Lagrangian space. In turn, the deformation of the survey window can also cause that regions are effectively unobserved in Lagrangian space, that were observed in Eulerian space (see Figure \ref{fig:winvuds} e.g. around redshift $z\sim 2$). As expected, we can see that the deformation is stronger towards lower redshifts, where the growth of structures is more evolved. 
\section{Large-Scale Galaxy Bias}
\label{sec:lsbias}

The \texttt{COSMIC BIRTH} algorithm  accounts for stochastic and non-linear Lagrangian bias, while non-local Eulerian bias is modelled through the displacement field connecting Eulerian to Lagrangian space \citep[see Section \ref{sec:birth} and for further details in][]{birth}. 
The only free parameter is the large-scale Eulerian bias  \citep{Kaiser1984}, which needs to be determined from observations or simulations.
We rely on detailed bias studies of highly star forming galaxies due to the nature of the surveys considered in this study  (see  Section \ref{sec:bias_studies}). 
\subsection{Bias Studies in Simulations and Observations}
\label{sec:bias_studies}

The various galaxy properties are in general correlated with their clustering behaviour, and hence, are indicators of how they trace the underlying dark matter density field.
More massive and luminous galaxies, such as luminous red galaxies (LRGs) for instance \citep[see e.g.][]{BOSS2016} show a strong clustering, tracing mainly the peaks of the density field \citep{Kitaura2015}. These galaxies are passively evolving, showing low stellar formation activity and old stellar populations. 

Apart from LRGs, star forming galaxies can be identified with photometric techniques  \citep[see e.g.][]{Daddi_2004} and emission line spectroscopy, frequently called emission line galaxies (ELGs) in literature. 
The galaxy bias of highly  star forming galaxies, such as {[O\,{\sevensize II}]}, {[O\,{\sevensize III}]}, H$\alpha$ detected galaxies \citep[see e.g.][]{Delubac2017,Kaasinen2017}, UV emitting Lyman-break galaxies \citep{Kollmeier2003} and Lyman-$\alpha$ emitters (LAEs, see e.g. \citet[][]{Kennicutt1998}) have been extensively studied in the literature.

Unlike LRGs that trace only the densest peaks of the density field, ELGs can populate the density field at nearly the whole range of overdensities. Therefore, their bias is close to unity at low redshifts \citep{Favole2016}. However, in this work we consider higher redshifts, when the dark matter field was less evolved, displaying weaker density perturbations. Hence, these types of galaxies display an increasing bias towards high redshifts  \citep{Okada2016, Guo2019}.

Studies based on numerical simulations have shown the bias as a function of redshift and star formation rate, comparing the galaxy and matter density fields \citep[see Table 1 in][]{Chiang2013}. Within the VUDS survey a bias measurement for the redshift range of $2\le z \le 5$ has been accomplished in \citet[][]{Durkalec2015,Durkalec2018} using two-point clustering analysis. Similar analysis has been performed for the FMOS-COSMOS survey, studying the projected correlation function and from there estimating the bias at a median redshift of $\bar{z}=1.58$ \citep{Kashino2017}.




\begin{figure}
    \centering
    \hspace*{-0.2cm}
    \includegraphics[width=0.49\textwidth]{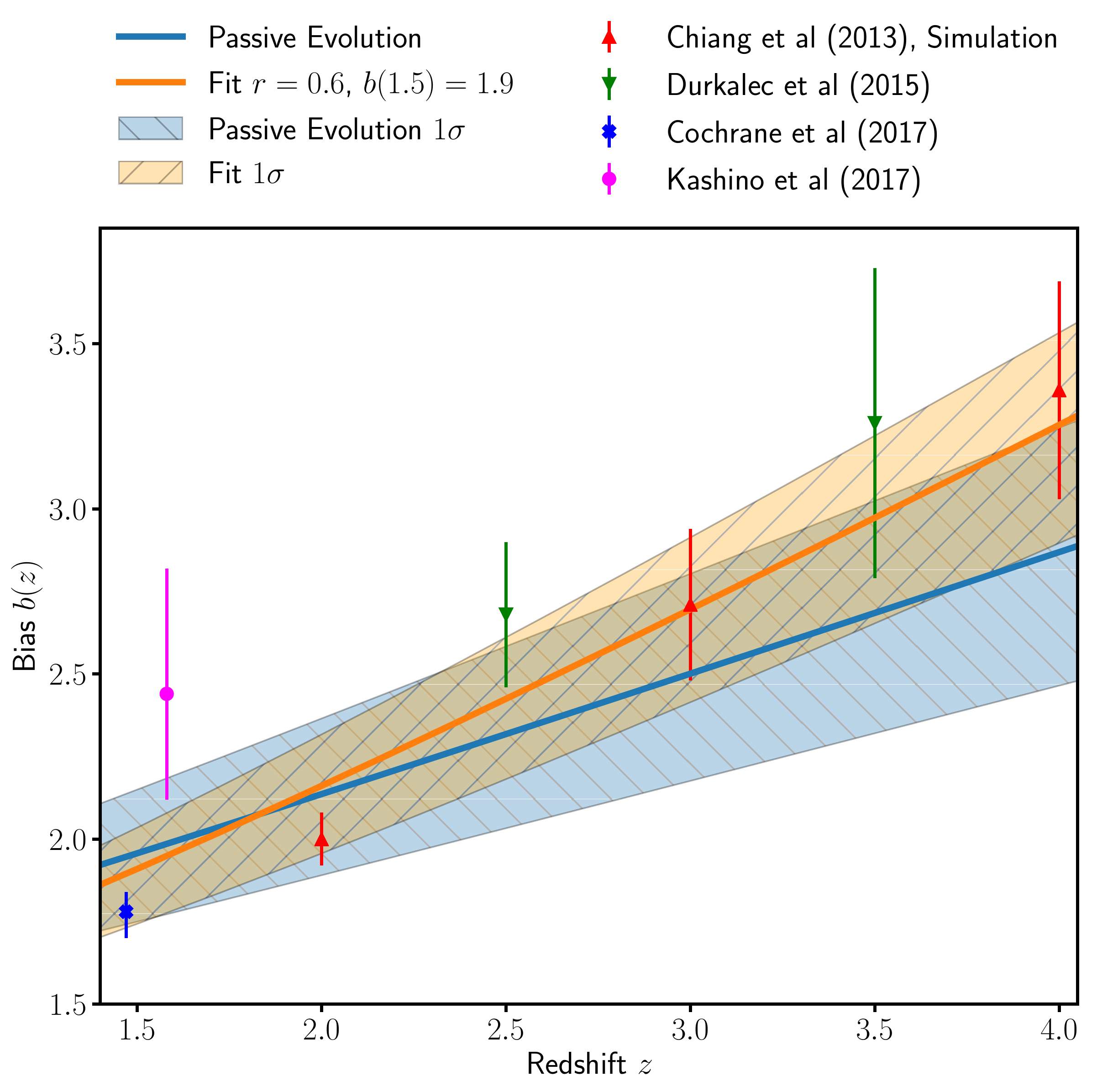}
    \caption{
    Large-scale galaxy bias $b(z)$ as a function of redshift $z$.
    The blue solid line and the corresponding blue shaded error band show the best fit for the passive evolution model of Equation \ref{eq:passive} with $b(z=1.5)=1.9\pm0.33$. The solid orange line and the orange shaded error band show the fitting result assuming Equation \ref{eq:redbias} with the parameters $r(z=1.5)=0.6$ and $b(z=1.5)=1.9\pm0.33$. The circles denote previous bias measurements whereas the red triangles show values derived from numerical simulations.}
\label{fig:bias_compare}
\end{figure}

\subsection{Large-Scale Bias Evolution}
\label{sec:lsbiasc}

In the following we explain our strategy to compute the large-scale bias, as required to perform the dark matter reconstructions throughout the redshift range of $1.4\leq z \leq 3.6$. 

Given that the galaxy populations from the 5 considered surveys are similar, we assume  as a null hypothesis, that galaxies share the same bias at a given redshift, and that their bias passively evolves through Equation \ref{eq:passive} (equivalent to a perfect correlation $r(z_2)=1$ in Equation \ref{eq:redbias}).
Then, we choose a narrow redshift range for which the deviation from passive evolution is expected to be negligible. In particular we select the range $1.4\lesssim z \lesssim 1.8$ embedded in a rectangular volume that extends from $2875\mperh\leq d_{\rm L} \leq  3387\mperh$ in line-of-sight distance. 
We perform a series of \texttt{COSMIC BIRTH} runs with the data set in this volume with a varying bias at $z=1.5$ from $1$ to $2.5$ (the general set-up of the \texttt{COSMIC BIRTH} reconstructions is described in Section \ref{sec:prep}). From these runs we find that the reconstructed primordial matter density shows unbiased power spectra w.r.t. the theoretical linear one for $b\approx1.9\pm0.3$  at redshift $z=1.5$ being conservative (and the rest of bias values given by passive evolution). This is in good agreement with the findings of \citet[][see Table 4, where they obtain  $b=1.78^{+0.06}_{-0.08}$ at $z=1.48$]{Cochrane2017}, and also roughly compatible with the clustering analysis of the FMOS-COSMOS survey  \citep[][see Table 3, finding $b=2.44^{0.38}_{0.32}$ at median redshift $\bar z = 1.59$]{Kashino2017} within the estimated uncertainties.
 
The passive evolution model \citep{Nusser1994,Fry1996}, based on our low redshift bias measurement, evolves to higher redshifts according to the blue solid line including the large associated uncertainties represented by the shaded light blue area in Figure  \ref{fig:bias_compare}.
From this figure we can assume that there is a trend in the bias measurements towards higher bias values with increasing redshift than what is predicted by passive evolution, although within the large uncertainties the passive evolution model is still nearly compatible with the data points.

In such a case the redshift evolution of the galaxy populations can be modelled. This can be achieved by introducing a correlation coefficient $r$ between the dark matter and the galaxy density field as a function of redshift \citep{Tegmark1998}.
As the selection criteria of the surveys do not significantly change with observed distance, we expect a moderate enhancement of the galaxy bias towards higher redshifts beyond the one described by passive evolution. Thus, we write:
\ba
\label{eq:redbias}
b(z_2) = \frac{\sqrt{\left(1-\frac{D(z_2)}{D(z_1)}\right)^2 - 2r(z_1) \left(1-\frac{D(z_2)}{D(z_1)}\right) b(z_1)  + b^2(z_1)}}{\left(\frac{D(z_2)}{D(z_1)}\right)} \,
\ea
for two redshifts $z_1$ and $z_2$ with $z_1<z_2$, a correlation coefficient $r(z_1)$, and the linear growth function $D(z)$ given by 
\ba
D(z) = \frac{H(z)}{H_0} \int\limits_z^\infty \frac{\rm{d}z^\prime}{H^3(z^\prime)}\bigg/ \int\limits_0^\infty \frac{\rm{d}z^\prime}{H^3(z^\prime)} \, ,
\ea
normalized to unity at redshift zero $D(z=0)=1$.
The evolution of the bias and the correlation coefficient are coupled as:
\ba
r(z_1) = \left(\left(1-\left(\frac{D(z_2)}{D(z_1)} \right) \right) +r(z_2)\left(\frac{D(z_2)}{D(z_1)} \right) b(z_2)\right)\bigg/b(z_1)  \,.
\label{eq:redcorr}
\ea
A perfect correlation of $r(z_2)=1$ at an earlier redshift $z_2$ will remain like that for all times, whereas $r(z_1)$ always tends towards $1$, regardless of its initial value, according to Equation \ref{eq:redcorr}.
In the case of a perfect correlation $r(z_2)=1$, Equation \ref{eq:redbias} equals Equation \ref{eq:passive} (described in Section \ref{sec:birth}) and no change of the galaxy population is expected along redshift. However, $r(z_2)<1$ implies a varying correlation coefficient, effectively describing a cosmic evolution of the galaxy distribution, which may be caused by galaxy formation or evolution  \citep{Tegmark1998}.

On the other hand, a series of \texttt{COSMIC BIRTH} runs disfavour $2\sigma$ deviations from the upper bias limits quoted in the literature, as they lead to unreasonable biased dark matter reconstructions (see Appendix \ref{app:bias_wrong}). We note, that the  selection function can lead to an excess of power in the two point statistics on large scales, and thereby higher bias values can be inferred \citep[see e.g.][]{2011PhRvL.106x1301T}.  
We therefore investigate the theoretical predictions for the bias evolution in simulations. As a result, we find that the galaxy samples considered in this work cover the stellar mass range of $M_\ast = 10^{9.5} M_{\sun}$ to $M_\ast = 10^{10.5} M_{\sun}$,  peaking at $ M_\ast \sim 10^{9.8} M_{\sun}$ (Lamaux et al. in prep.). According to this finding, we obtain the data points represented in red upwards pointing triangles in Figure \ref{fig:bias_compare} \citep[see Table 1 in][]{Chiang2013}.   
We find that these simulation based data are in agreement with the observational measurements, however favour slightly lower bias values.
In the spirit of being conservative, and avoid bias ranges which can be affected by selection effects, we include the simulation \citep[][]{Chiang2013} and observational \citep[][]{Cochrane2017,Kashino2017,Durkalec2015,Durkalec2018} data points in a least squares fit. The resulting  bias evolution model is represented in solid red with the uncertainty given by the light red shaded area in Figure \ref{fig:bias_compare}.

According to this bias study, we have found some moderate evidence (given the hitherto large uncertainties due to the small volumes covered by high redshift galaxy surveys) for a cosmic evolution of the galaxy bias beyond passive evolution, hinting towards ongoing galaxy formation and merging processes at these redshifts.
A coefficient of $r(z=1.5)=0.6$ corresponds to $r(z=3.6)=0.42$ (see Equation \ref{eq:redcorr}). This apparently tiny variation, reduces the tension with observationally constrained biases at redshift $z>1.5$.
However, a proper  verification of a deviation from passive evolution requires a deeper study, extending the runs we performed at low redshift to higher ones. The available data at this stage might not be sufficient to make stronger claims and we leave a more detailed investigation of the bias evolution to a forthcoming work.

The resulting large-scale bias calculations presented in this section can be fed into the \texttt{COSMIC BIRTH} code to produce unbiased dark matter reconstructions, as shown in Section \ref{sec:reconstructions}.

\section{\texttt{COSMIC BIRTH} Applied to the COSMOS Field}
\label{sec:reconstructions}
In this section we present the application of \texttt{COSMIC BIRTH} code \citep[see][and Section \ref{sec:birth}]{birth} to the spectroscopic surveys in the COSMOS field (see Section \ref{sec:surveys}). Previous pioneering COSMOS density field estimates were based on photometric redshifts \citep{Kovac2009} and used tessellations of the observed galaxy fields \citep{Scoville2013,Darvish2015}, but did not account for the selection function \citep{2012MNRAS.424..553A,Smolcic2017}, or focused on individual high density peaks \citep[e.g.][]{Wang2016}. Therefore, this work represents the first comprehensive study to address all the above mentioned issues, taking into account structure formation, selection functions, redshift-dependent bias descriptions and redshift-space distortions within a forward Bayesian analysis.

\subsection{Setup of the Reconstructions}
\label{sec:prep}

The \texttt{COSMIC BIRTH} code in its current version performs calculations on  cubical regular meshes in comoving Cartesian coordinates and uses the corresponding galaxy positions in redshift-space on the light-cone with their corresponding bias obtained from galaxy catalogues as input source.

First we assume in this study a $\Lambda$CDM model with cosmological parameters defined in Equation \ref{eq:cosm}. 
Then we select the input catalogues comprehending 5 different redshift surveys, as described in Section \ref{sec:survey2}. The large-scale bias as a function of redshift is obtained from the data itself and further constrained according to some previous studies, as explained in Section \ref{sec:lsbiasc}. From this the connection to Lagrangian bias including a nonlinear and nonlocal treatment is internally computed, as explained in Section \ref{sec:bias}. With the given cosmology we can translate the angular and redshift coordinates for each galaxy into comoving Cartesian coordinates  $(x,y,z)$.
The corresponding geometry and angular completeness to each survey is computed, as explained in detail in Section \ref{sec:angular}.
While the \texttt{COSMIC BIRTH} code does not assume the plane parallel approximation at any step, this approximation is nearly fulfilled given the large distances to the galaxies of the considered redshift range and the narrow angular coverage (see Figure \ref{fig:nz}).
We take advantage of that for visualisation purposes, and choose a coordinate transformation so that the centre of the zCOSMOS-deep survey is aligned to $\mathrm{DEC} = 0^\circ$ and $\mathrm{R.A.} = 180^\circ$, which makes the $Y$-axis approximately coincide with the declination angle (for angles close  to zero, as in this case) and the $X$-axis with the redshift $z$ (see e.g. Figures \ref{fig:winvuds},\ref{fig:dens4box}). 
The final results are presented in the original coordinate system. 
The reconstructions comprise the galaxies of the five mentioned surveys within a redshift range of $1.4 \leq z \leq 3.6$, which translates into a comoving distance of $d_{\rm Box} = 1898 \mperh$ along the line-of-sight. 
For computational reasons we have split the reconstruction into four cubical volumes making sure that a large enough volume (in terms of mode-coupling) is taken in each case of $512 \mperh$ side length \citep{Sorce2016}.
This resulted in a mesh grid resolution resolution of $2 \mperh$ using meshes of $256^3$ voxels. We placed the four volumes successively along the line-of-sight considering an overlapping region of $50 \mperh$ on each side. This is an adequate choice, acknowledging that the galaxy-galaxy correlation function drops steeply at scales larger than 20 $\mperh$ \citep[see e.g.][]{Anderson_2012}. However, boundary effects (e.g. velocity correlations) at the transition region of the sub-volumes are the primary source of uncertainty in the reconstructions, which will be further analyzed in forthcoming works. Additionally, we did not place galaxies in a buffering zone of $25 \mperh$ at the edges of each volume, which we took into account within the radial selection function accordingly. This means, that the line-of-sight data region is 5-7 times larger than in transverse directions for each sub-volume (see Figure \ref{fig:scatter_comb}).
Thus, the full reconstructions extends from $2875 \mperh$ to $4773 \mperh$ in comoving line-of-sight distance. We note, that
\texttt{COSMIC BIRTH} code takes light-cone evolution into account within each reconstructed volume. We choose 8 redshift bins for each sub-volume \citep[see Figure 4 in][]{birth}.

\subsection{Numerical Assessment \& Convergence}
\label{sec:tests}

\begin{figure}
    \centering
    \hspace*{-0.5cm}
    \includegraphics[width=0.49\textwidth]{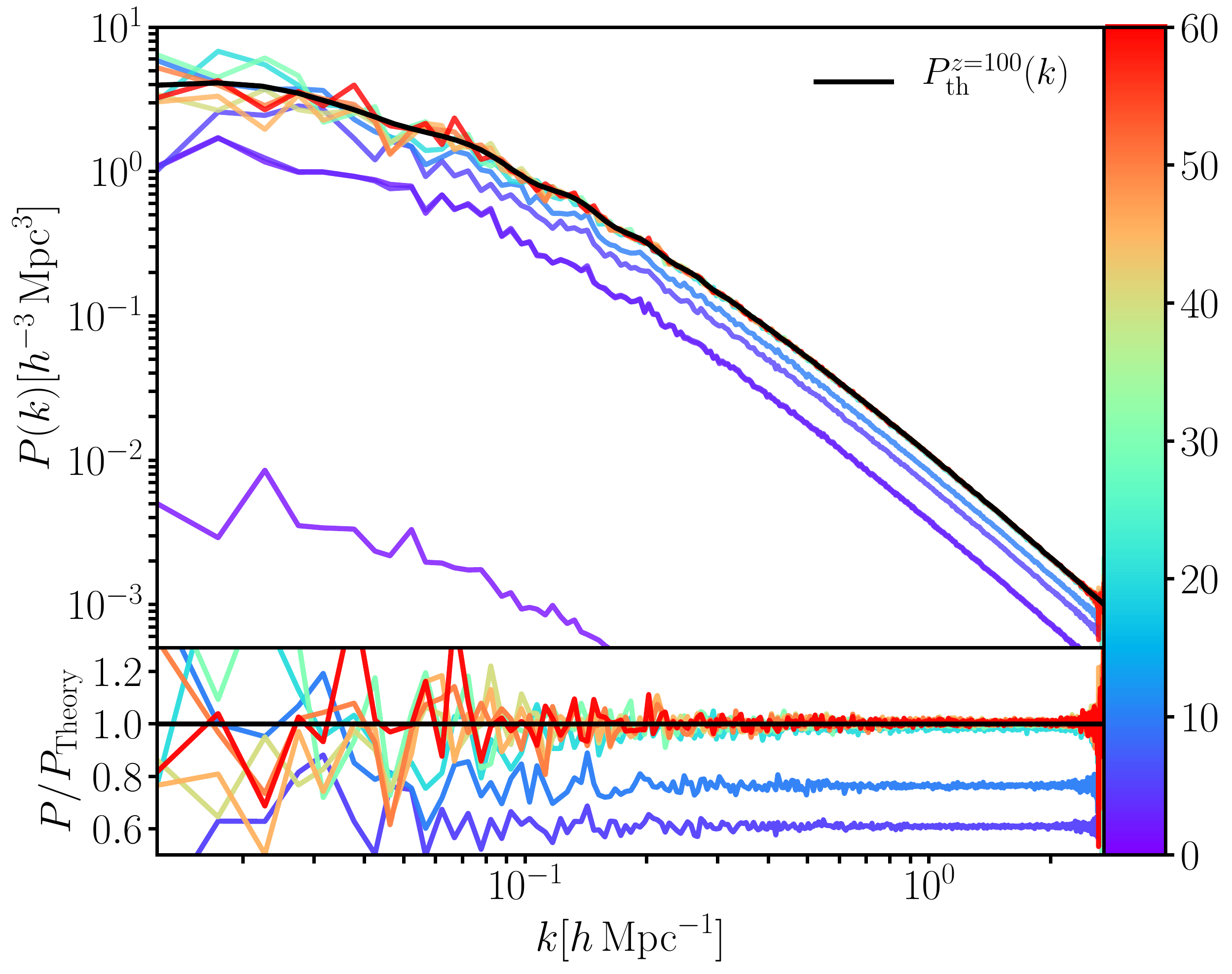}
    \caption{Power spectra of the inferred initial density field $\delta(q)$ at $z=100$ shown colour coded for the first 60 HMC samples for the sub-volume ranging from $1.7 \leq z \leq 2.2$. In the bottom panel the ratio of the power spectra and the theoretical prediction are shown. Convergence is roughly achieved after the $\sim$ 40th samples.}
    \label{fig:ps_convergence}
\end{figure}
\begin{figure}
    \centering
    \hspace*{-0.5cm}
    \includegraphics[width=0.49\textwidth]{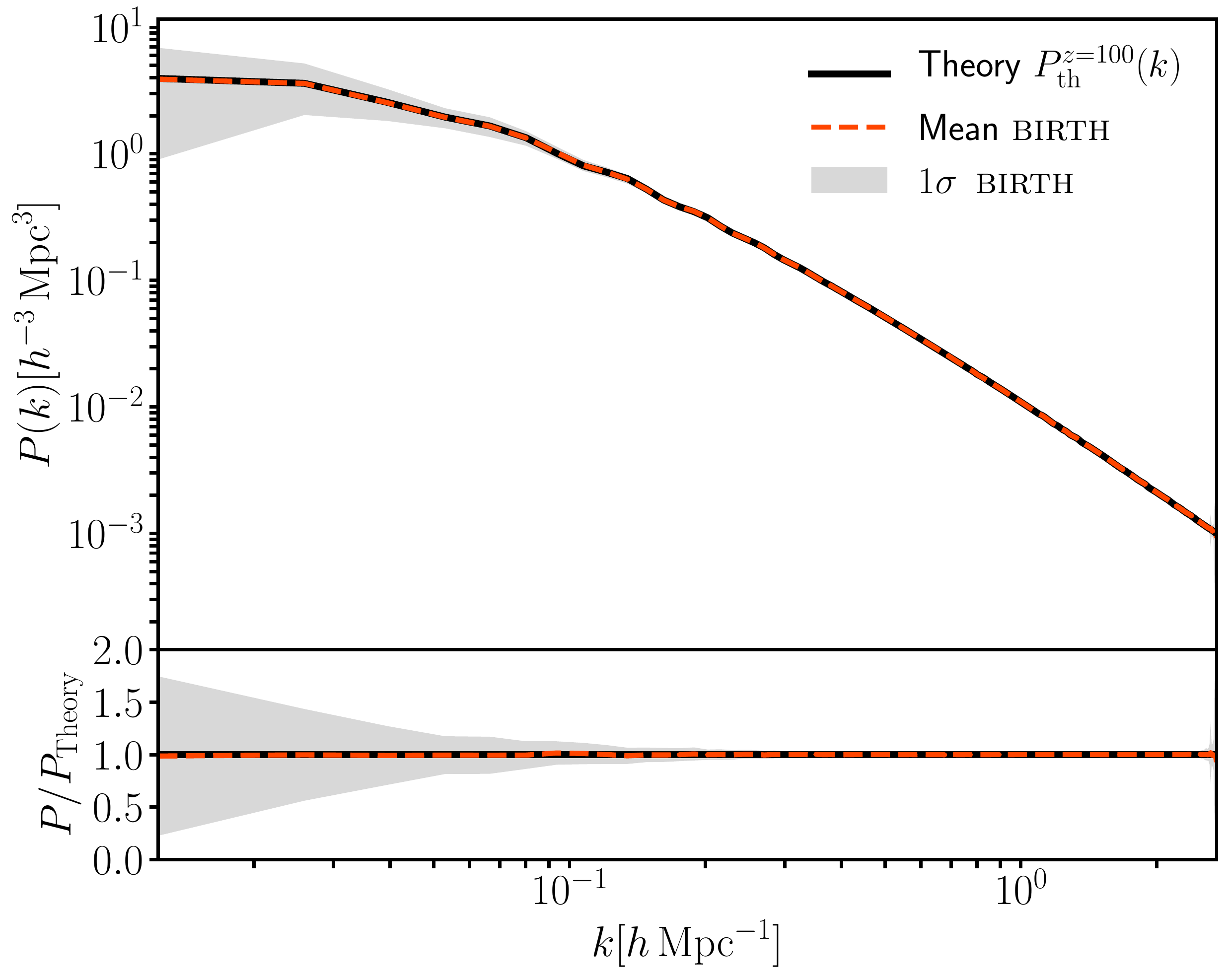}
    \caption{Mean power spectrum of the inferred initial density field $\delta(q)$ shown for the sub-volume ranging from $1.7 \leq z \leq 2.2$. The red dashed line represents the mean power spectrum averaged over 6000 individual HMC realisations with $1\sigma$ standard deviation shown as a grey band. In the bottom panel we show the ratio of the mean power spectrum and its uncertainty band with the theoretical $\Lambda$CDM prediction.}
    \label{fig:ps_mean_data}
\end{figure}

Our numerical tests showed that the \texttt{COSMIC BIRTH} reconstructions are insensitive in terms of the power spectrum to deviations from up to 30\% from our best large-scale bias evolution estimate shown in Figure \ref{fig:bias_compare}. We note, that this already excludes the high end of bias values allowed within $2\sigma$ confidence levels by the observations (see discussion in  Section \ref{sec:lsbiasc}).
The  convergence behaviour of the power spectra from the Lagrangian density fields (for the best large-scale bias model), starting from a perfectly homogeneous density field $\mbi\delta(\mbi q)=0$ is shown in Figure \ref{fig:ps_convergence}.
The reconstructions show unbiased power spectra already after the 40th iteration step. Up to this iteration we consider the chain to be in the burn-in phase and not representing the target distribution of which we aim at drawing samples from. 
Furthermore, the top panel of Figure \ref{fig:ps_mean_data} shows the mean power spectrum of the initial density field $\delta(q)$ at $z=100$ averaged over 6000 samples, illustrating the $1\sigma$ standard deviation as a grey band. The bottom panel shows the corresponding ratios with the theoretical linear $\Lambda$CDM prediction. The reconstructed density fields show unbiased power spectra over all scales, confirming an accurate bias treatment. 
The effect of a mishandled bias in our study is shown in Appendix \ref{app:bias_wrong}. We note that sampling the large-scale bias from sparse data sets as considered here with such a low data volume fraction is going to be inaccurate, although the Bayesian framework allows for it \citep[see][]{Granettvimos,Jasche2017}.
\begin{figure}
    \centering
    \hspace*{-0.3cm}
    \includegraphics[width=0.49\textwidth]{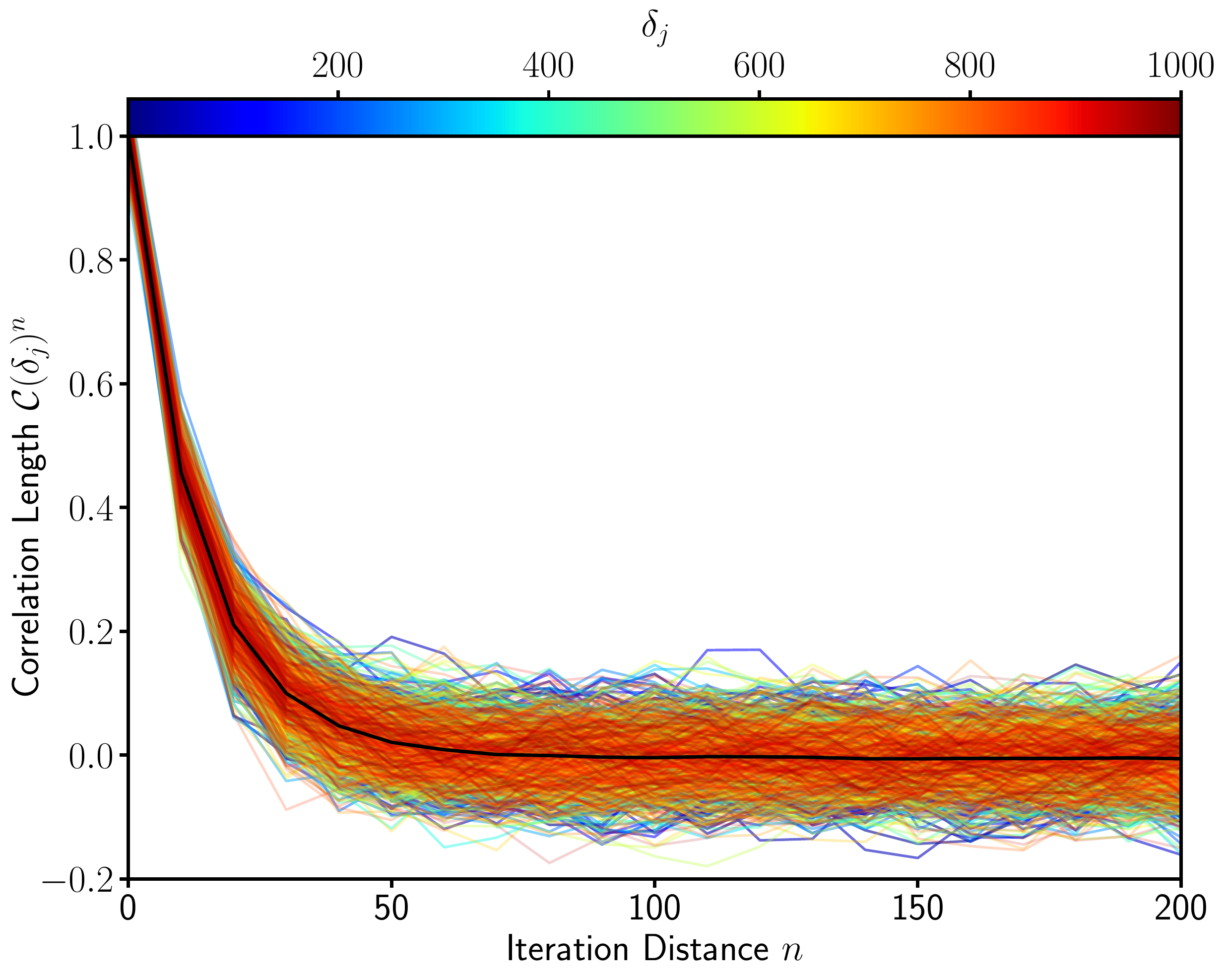}
    \caption{Correlation length $\mathcal{C}(\delta_j)^n$ of 1000 randomly chosen density field voxels $\delta_j$, where the completeness is $w_j>0$. We compute the correlation length over a sample of $N=6500$ HMC realization with an iteration length $n\in[0...200]$ shown for the sub-volume ranging from $1.7 \leq z \leq 2.2$. The colour code indicates the correlation length for a particular $\delta_j$ while the black solid line represents the average over all density voxels. We consider $\mathcal{C}(\delta_j)^n<0.1$ to be uncorrelated, which our Gibbs-sampling chain drops below after $n=40$ iterations. Thus, each 40th  realization is an independent sample.}
    \label{fig:corr}
\end{figure}
\begin{figure}
    \centering
    \hspace*{-0.5cm}
    \includegraphics[width=0.51\textwidth]{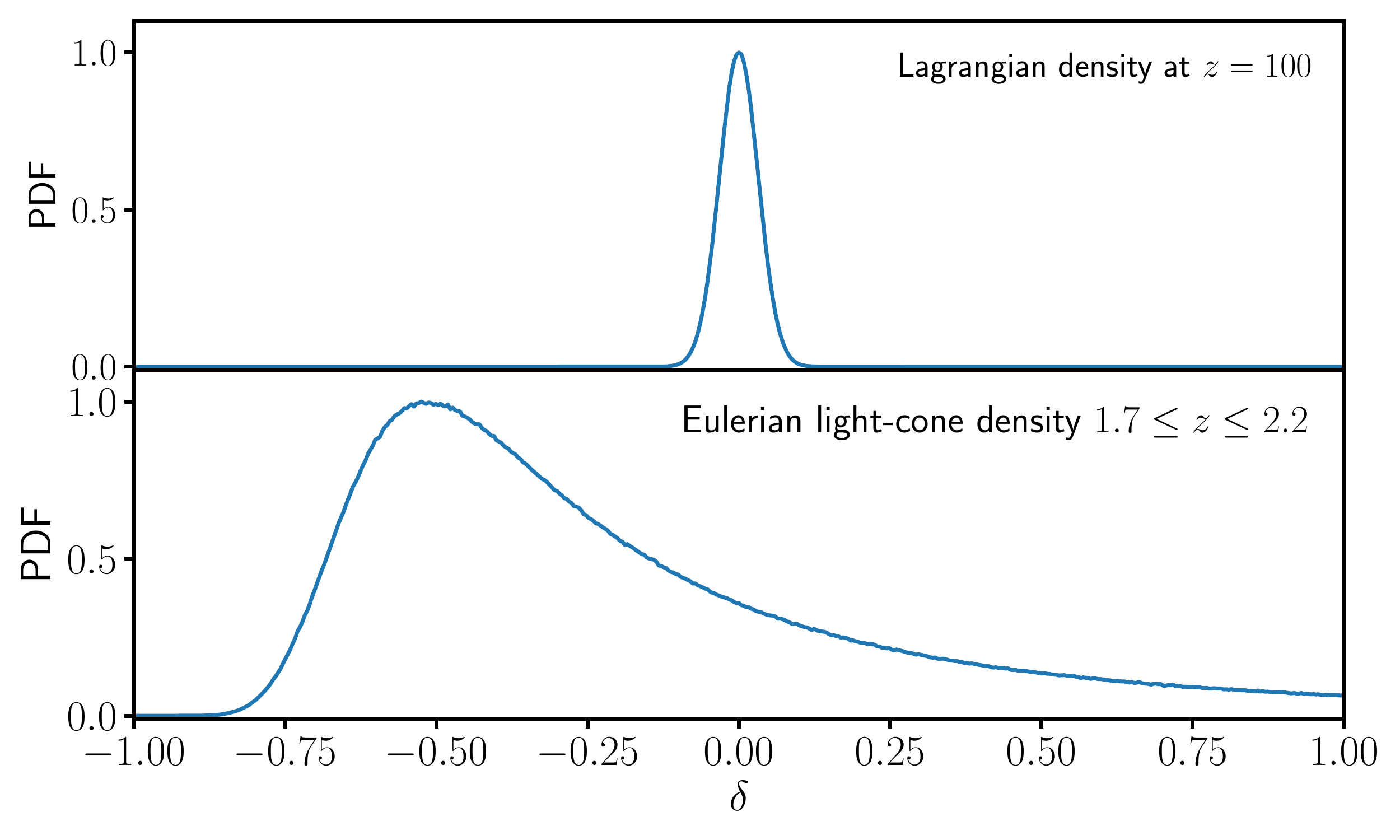}
    \caption{Histogram of the inferred density fields normalized to unity for $-1 \leq \delta \leq 1$. {\bf The top panel} shows a histogram of the density at initial conditions (Lagrangian coordinates) at $z=100$. {\bf The bottom panel} shows a histogram of the light-cone density field in Eulerian coordinates in the range of $1.7 \leq z \leq 2.2$. The Lagrangian density on the top panel shows a closely Gaussian distribution with a negligible skewness of $s = 0.001$, while the Eulerian density on the bottom panel presents a highly skewed distribution with $s = 5.373$.}
    \label{fig:dens_pdf}
\end{figure}

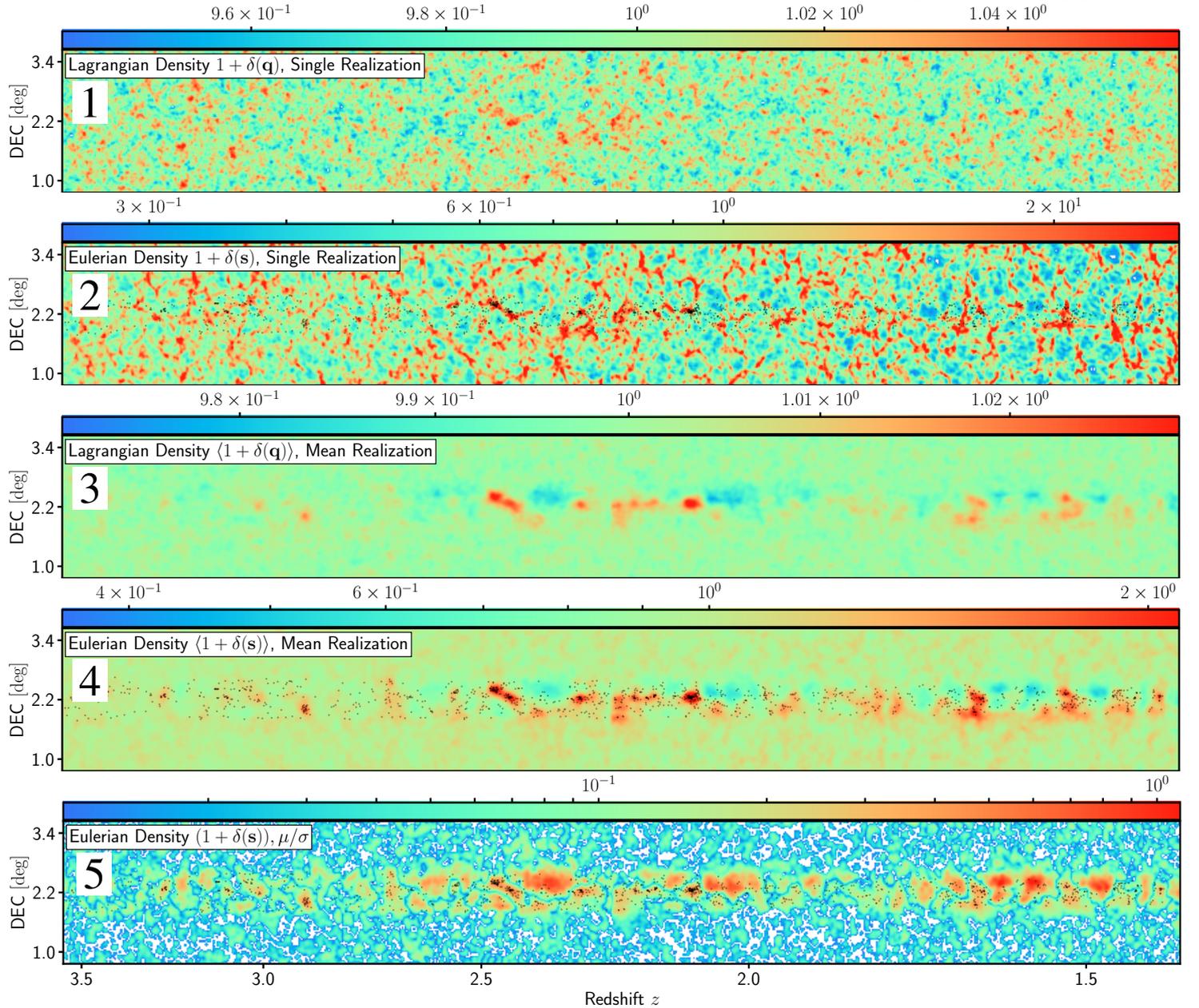
\begin{figure*}
    \centering 
    \begin{picture}(100,100)
    \put(-240,0){    \includegraphics[width=1.1\textwidth]{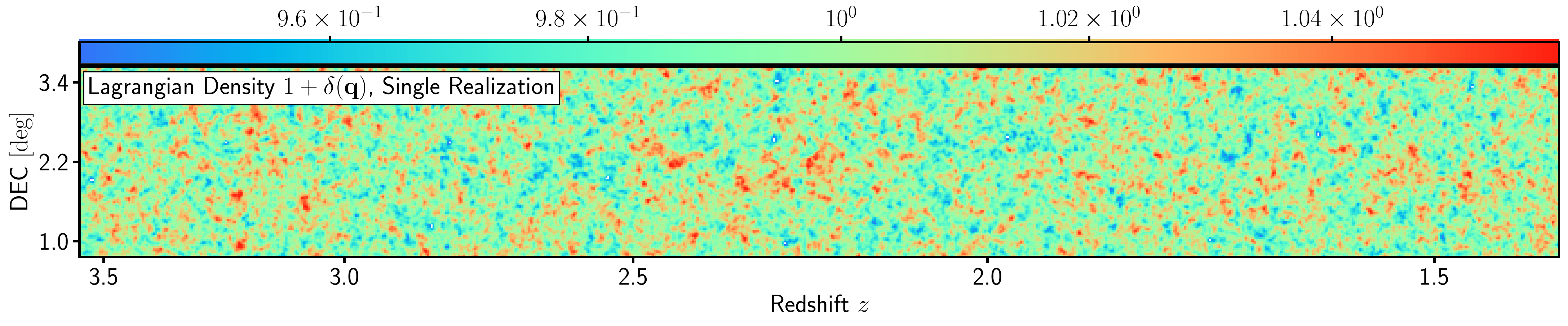}
    }
    \put(-205,60){\colorbox{white}{\Huge 1} }
    \end{picture}

    \begin{picture}(100,100)
    \put(-240,10){    \includegraphics[width=1.1\textwidth]{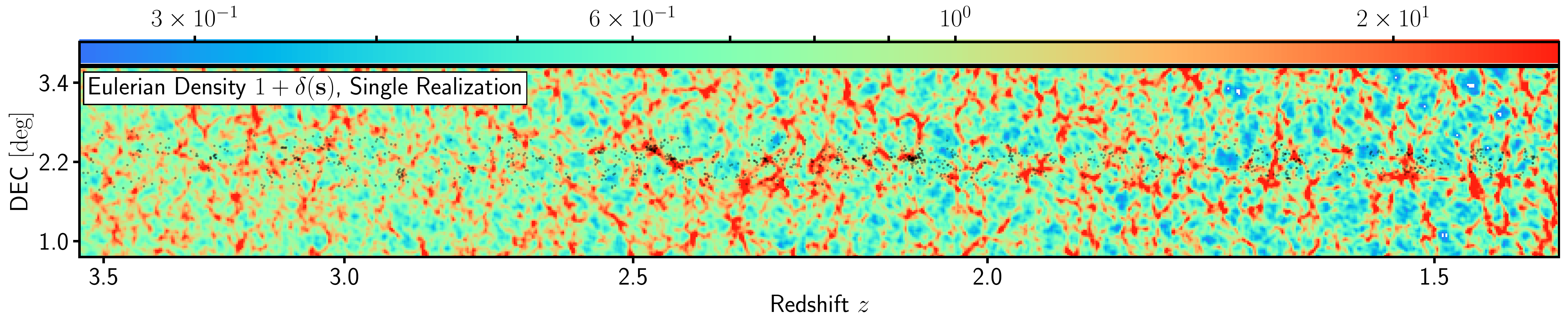}
    }
    \put(-205,70){\colorbox{white}{\Huge 2} }
    \end{picture}
    
    \begin{picture}(100,100)
    \put(-240,20){    \includegraphics[width=1.1\textwidth]{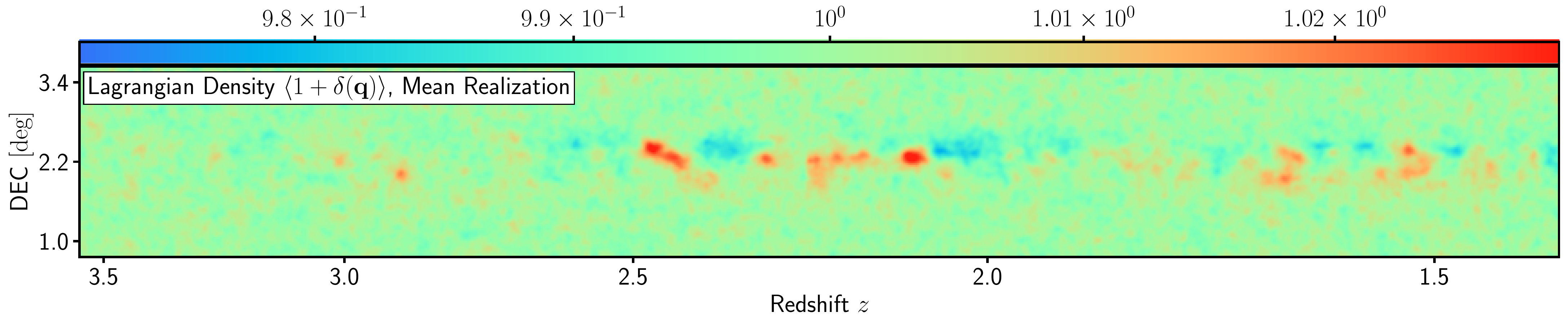}
    }
    \put(-205,80){\colorbox{white}{\Huge 3} }
    \end{picture}
    
    \begin{picture}(100,100)
    \put(-240,30){    \includegraphics[width=1.1\textwidth]{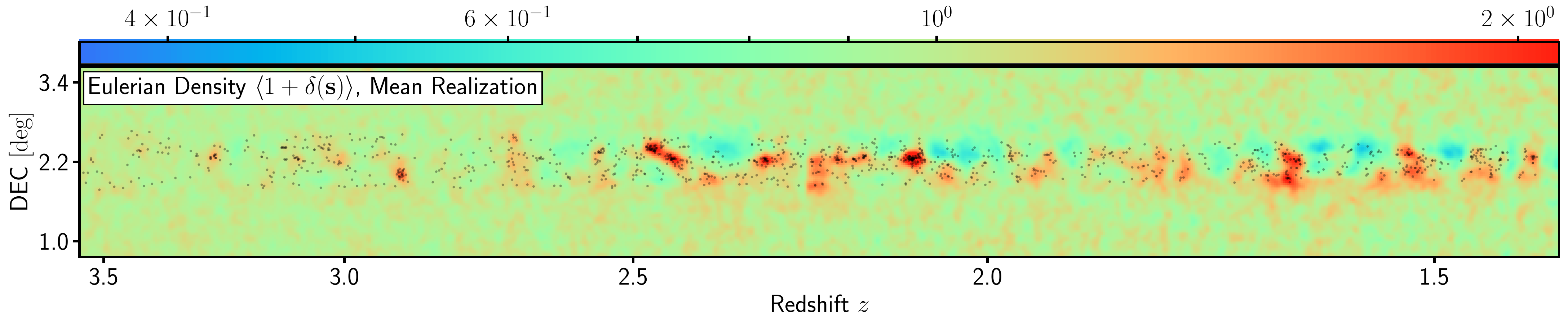}
    }
    \put(-205,90){\colorbox{white}{\Huge 4} }
    \end{picture}
    
    \begin{picture}(100,100)
    \put(-238.5,40){    \includegraphics[width=1.1\textwidth]{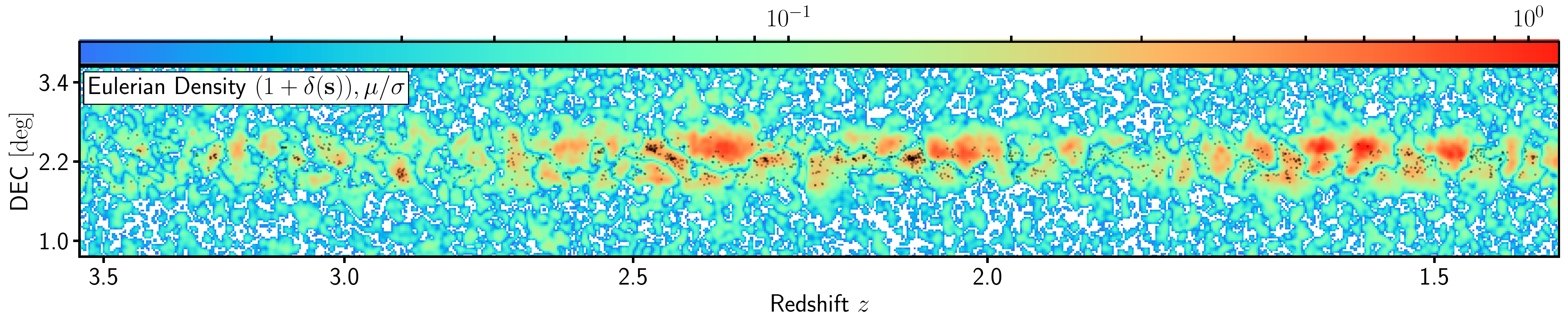}
    }
    \put(-202,100){\colorbox{white}{\Huge 5} }
    \end{picture}
    \vspace*{-1cm}
    \caption{\texttt{COSMIC BIRTH} reconstructed density field $1+\mbi \delta$ shown as slices in line-of-sight and declination coordinates with thickness of $6 \mperh$. Phase-space mapping with the corresponding tetra-hedra tesselation has been performed as mentioned in Section \ref{sec:birth}. The whole declination angle on the $Y$-axis corresponds to $200 \mperh$ in comoving coordinates. From top to bottom the two upper slice plots, panels 1 and 2, show firstly a single realization the Lagrangian initial density field $1+\delta(\mbi q)$ at $z=100$ and secondly a light-cone realization at Eulerian redshift-space $1+\mbi\delta(\mbi s)$ with the individual galaxy positions plotted on top. Panels 3 and 4 represent the mean distributions of the inferred initial Lagrangian $\langle 1+\mbi\delta(\mbi q)\rangle$ at $z=100$ and final Eulerian $\langle 1+\mbi\delta(\mbi s)\rangle$ density fields averaged over 6000 HMC realizations, respectively, where on top of the Eulerian density field the galaxy positions are plotted with black dots. Finally, we show in panel 5 the signal-to-noise ratio $\mu/\sigma$ of the Eulerian density field with the galaxy positions plotted on top.}
\label{fig:dens4box}
\end{figure*}

\begin{figure*}
    \hspace*{-0.75cm}
    \includegraphics[width=1.1\textwidth]{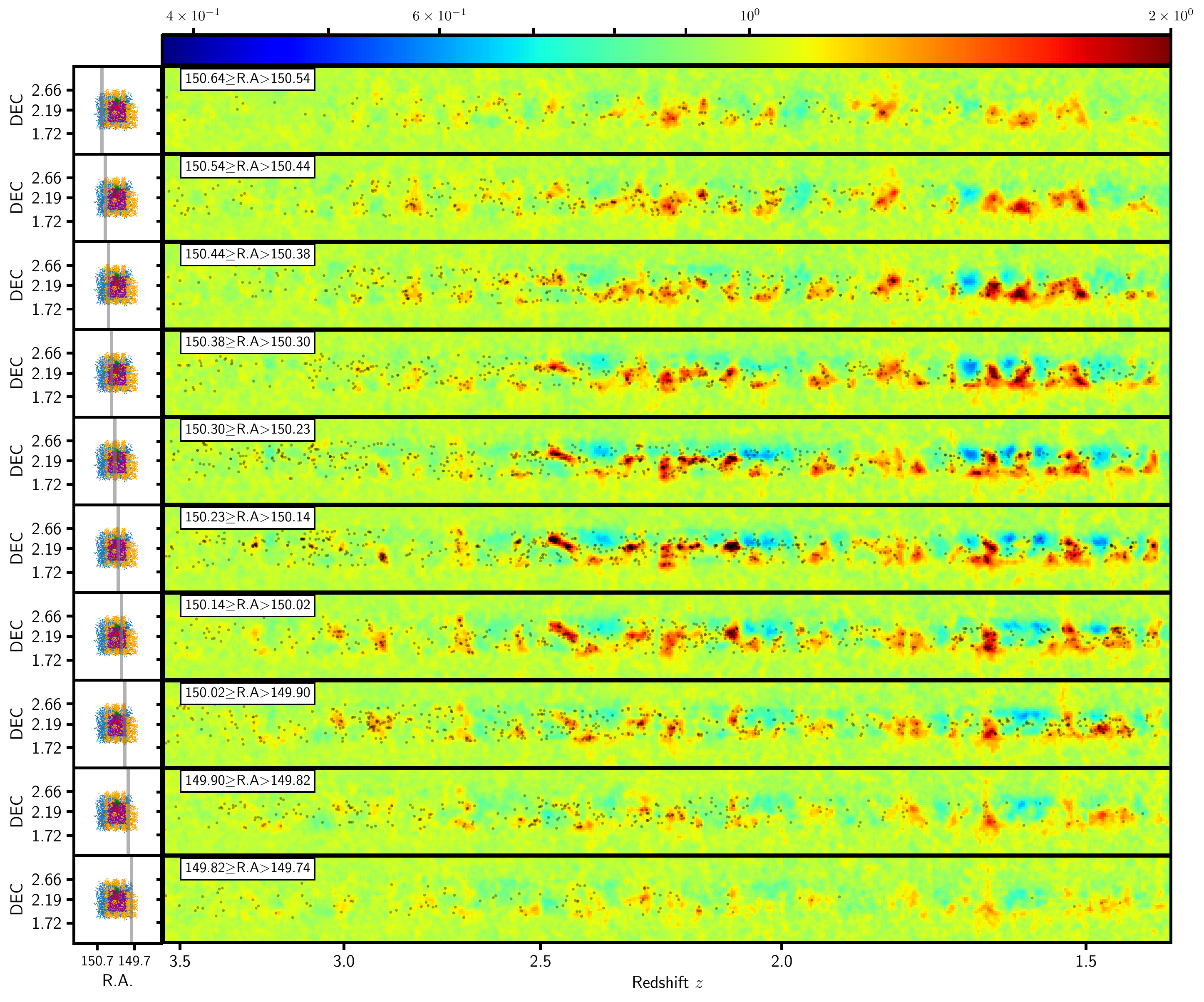}
    \caption{Tomographic slices of the reconstructed matter density field in the declination-redshift plane. On the left we show the angular footprint of the 5 surveys in the same colour code as Figure \ref{fig:scatter_comb} with a grey line showing the R.A. thickness of the corresponding slice. On the right the mean light-cone density reconstructions are shown.}
\label{fig:ra_sliced}
\end{figure*}

\begin{figure*}
    \centering
    \hspace*{-1cm} \vspace*{-0.2cm}
    \includegraphics[width=1.025\textwidth]{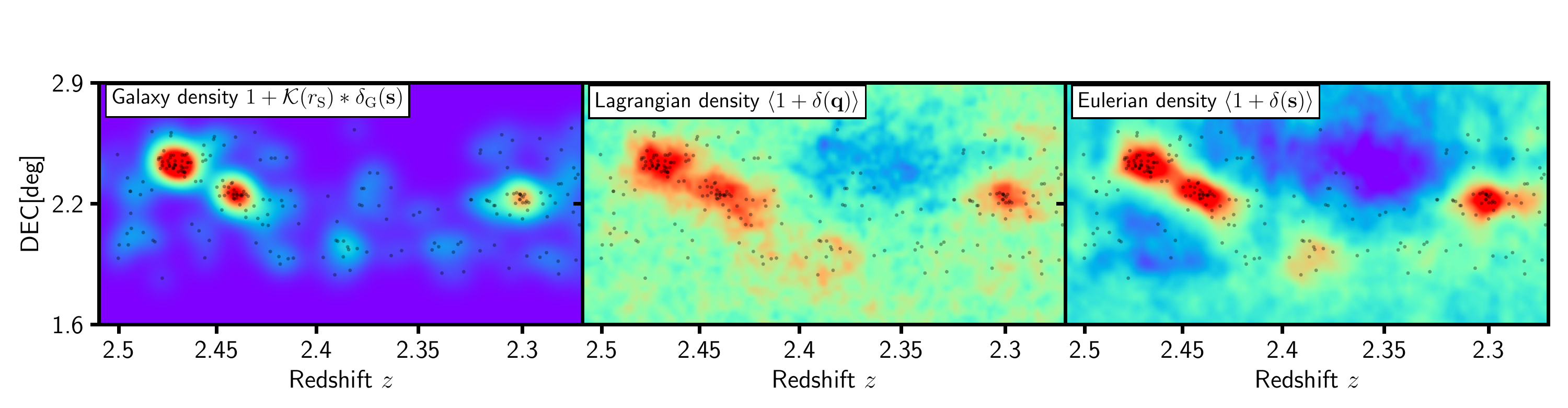}
    \caption{Density plots for a zoomed region. On the left panel the galaxy density $1+\mathcal{K}(r_{\rm S})\ast \delta_{\rm G}(\mbi s)$ is shown convoluted with a Gaussian smoothing kernel $\mathcal{K}(r_{\rm S})$ with smoothing radius of $r_{\rm S}=2\mperh$, followed by the mean density field in initial (middle panel) $1+\delta(\mbi q)$ and final (right panel) conditions $1+\delta(\mbi s)$ for the same slice as shown in Figure \ref{fig:dens4box}. Overplotted with black dots are shown the corresponding galaxy positions.}
\label{fig:compare}
\end{figure*}
Once the cosmology has been chosen, and the data input defined, there is only one free parameter in the \texttt{COSMIC BIRTH} code, which is the large-scale bias of the different populations. As shown in Section \ref{sec:lsbias}, there is a confidence region for the large-scale bias as a function of redshift for the galaxies considered in this study. Given the the low volume filling fraction of the data region with respect to the entire volume, which needs to be considered to keep the mode-coupling effects from large-scale modes low \citep[see e.g.][]{Sato2009,Takada2013}, the question arises whether the reconstructions are sensitive to bias. 
To validate the independence of the HMC samples, we calculate the correlation length for the inferred density field. The correlation length $\mathcal{C}(\delta_j)^n$ for a particular density voxel $\delta_j$ at an iteration distance $n$ over $N$ samples is given by:
\ba
\mathcal{C}(\delta_j)^n = \frac{1}{N-n} \sum\limits_{i=0}^{N-n} \frac{(\delta_j^i-\langle\delta_j\rangle)}{\sqrt{\sigma^2(\delta_j)}}\frac{(\delta_j^{i+n}-\langle\delta_j\rangle)}{\sqrt{\sigma^2(\delta_j)}} \, ,
\ea
where $\langle\delta_j\rangle = \frac{1}{N}\sum_i \delta_j^i$ is the mean of the density voxel $\delta_j$ over $N$ samples and $\sigma^2 (\delta_j) =  \frac{1}{N}\sum_i \left( \delta_j^i-\langle\delta_j\rangle \right)^2 $ the corresponding variance.
We show the correlation length in Figure \ref{fig:corr} for 1000 randomly chosen density voxels $\delta_j$, with $j\in [1...1000]$ in the data region of our reconstructed volume. This demonstrates that we draw independent samples each $\sim$40th Gibbs-sampling iteration.

To quantify the Lagrangian to Eulerian mapping using ALPT within \texttt{COSMIC BIRTH}, we show the density distribution of the initial and final density fields in Figure \ref{fig:dens_pdf}. On the same scale on the x-axis the density field value $\delta$ is plotted against the normalized probability distribution. We can see that the density field in Lagrangian space follows closely a Gaussian distribution with mean peaked at zero, and very small variance and skewness, $ \mu \approx 0, \sigma^2  = 0.001, s = 0.001$. The histogram for the Eulerian density field also has a mean value close to zero, $\mu \approx 0$, however shows a variance of $ \sigma^2 = 0.95$, and a skewness of $s = 5.373$. 
This is induced by gravity over cosmic time scales and is in excellent agreement to previous findings studied in detail in \citet[][]{Neyrinck2013}, comparing the displacement fields of second order Lagrangian perturbation theory (2LPT) and $N$-body simulations at different redshifts. Especially at redshifts $z\geq1$ the 2LPT displacements and the $N$-body results are in very good agreement and thus represent a very reasonable choice for this study (even more so, since we use ALPT).  
\subsection{Density Inference Results}
The resulting reconstructions corresponding to our best large-scale bias evolution estimation are shown in Figure \ref{fig:dens4box} as slice plots. On the $X$-axis we show the line-of-sight distance\footnote{We note that the line-of-sight distance corresponds to the redshift of the observations only for the light-cones. For the initial density fields $\mbi\delta(\mbi q)$ at $z=100$, $X$-axis only shows a distance measure.} as redshift $z$, and on the $Y$-axis we show the declination ($\rm DEC$). The slices are shown with a thickness of $6\mperh$ (averaging over three neighbouring cells). The corresponding convergence behaviour, power spectra, and matter statistics was discussed in Section \ref{sec:tests} and shown in Figures \ref{fig:ps_convergence}--\ref{fig:dens_pdf}.

We find  homogeneously  distributed initial cosmic density fields (see first panel of Figure \ref{fig:dens4box} for individual  reconstructions of $1+\mbi\delta(\mbi q)$). This is further demonstrated in the statistical matter distribution shown in the upper panel of  Figure \ref{fig:dens_pdf}) with negligible skewness and kurtosis values, as expected for a Gaussian density field. Furthermore, the top panel of Figure \ref{fig:dens4box} shows no redshift evolution.
This is in contrast with the second upper panel, in which the Eulerian density for the corresponding reconstruction $1+\mbi\delta(\mbi s)$ is shown.
In fact, the corresponding matter statistics shown in the lower panel of Figure \ref{fig:dens_pdf} shows a highly non-Gaussian distribution.
Further inspection of the second panel of Figure \ref{fig:dens4box}, shows an increase of the matter fluctuations towards low redshifts, particularly enhanced through the appearance of large cosmic voids, depicted in dark blue. The corresponding ensemble averages  (over 6000 Gibbs-sampling iterations) to panels 1 and 2 are shown in the  panels 3 and 4, $\langle 1+\mbi\delta(\mbi q) \rangle$ and $\langle 1+\mbi\delta(\mbi s) \rangle$, respectively.
In particular, the average over the ensemble of Eulerian density fields (i.e., the expected dark matter field from a Bayesian calculation) shows a high correlation with the galaxy field, as it should happen, when the initial cosmic density field is accurately recovered.
One can also appreciate the vanishing fluctuations in regions with low completeness, in concordance  with a Bayesian analysis (see for a comparison, as an example the completeness of VUDS depicted in Figure \ref{fig:winvuds}).
Panel 5 shows the signal-to-noise ratio obtained through the ratio of the mean density over its standard deviation $\mu/\sigma$. As expected, we find a higher signal-to-noise ratio in the data region. Interestingly, cosmic voids are particularly prominent in this measure. The missing data or equivalently, empty window function regions increase the uncertainty of the corresponding density cells and thus the variance of the density field (see Equation \ref{eq:expect}). This would be reflected in the signal-to-noise ratio, shown in the bottom panel of Figure 11. We however see no significant decrease in the signal-to-noise ratio at the gaps of the VUDS survey.

To further assess the matter distribution in the reconstructions, we show tomographic slice plots for various R.A. ranges (grey vertical stripes) in Figure \ref{fig:ra_sliced}, showing the survey footprints on the left-hand side and the corresponding density reconstructions next to it on the right. This permits us to evaluate the extension of the proto-clusters and cosmic voids. 
A comparison between the Lagrangian and Eulerian ensemble average reconstructions (middle panels in Figure \ref{fig:dens4box})  shows higher density fluctuations in Eulerian space, and small displacements of the center of mass of the proto-clusters towards high redshifts. Focusing on the lowest redshift regions one can observe stronger variations of the shape of the proto-clusters. 

To further investigate this, we zoom into a prominent proto-cluster region with $2.28 \leq z \leq 2.51$ and $1.6^\circ\leq \mathrm{DEC} \leq 2.9^\circ$ (see Figure \ref{fig:compare}).
This comparison shows three high density proto-clusters at $z\sim 2.45$, $z\sim 2.4$ and $z\sim 2.3$  growing  throughout the cosmological timescale starting from $z=100$ down to their light-cone redshifts.
At first glance, the proto-clusters seem to grow in place, with negligible displacements, in accordance with linear perturbation theory, describing the growth of perturbation fixed in comoving frame between the initial redshift $z_q$ to the final redshift $z_f$ with $\delta(\mbi r,z_f) = \delta(\mbi r,z_q)\; \nicefrac{D(z_f)}{D(z_q)}$, where $D(z)$ is the linear growth factor. Linear theory from $z_q=100$ to the redshift of our observed light-cone ($\sim z_f=2.3$) predicts a growth by  a factor of approximately $30$, which is consistent with the colour bar of the first and second panel of Figure \ref{fig:dens4box}.
However, a more detailed inspection reveals a complex non-spherical accumulation of mass when comparing the middle and the right panels. This can be well appreciated when comparing to the Gaussian smoothed galaxy field shown on the left panel. We also find  from this comparison how the two proto-clusters on the left, which appear as separate entities in the left panel, are actually connected, most likely having a dark matter bridge in between their respective galaxy distributions. It is also interesting to see the accumulation of matter through the action of gravity from a ring-like shape cloud to a spherical overdensity region in the region centred at $z\sim 2.38$ and $\mathrm{DEC}\sim 1.9^\circ$, comparing the Lagrangian and Eulerian density fields.  We also focus on the prominent $z\sim2.1$ proto-cluster and show how this region is reconstructed within different individual realizations in Appendix \ref{app:cluster}.
This Bayesian analysis taking the completeness into account, also reveals  regions that are more likely to be true cosmic voids. In particular, the right panel shows a deep void towards higher declination angles in the redshift range $2.3<z<2.4$, which feeds the overdensity peaks on its left and right. 
This is in agreement with the cosmic void initially discovered from the three dimensional Ly$\alpha$ forest tomography \citep[][]{Krolewski2018}. Another void region is forming directly under the proto-clusters at $z\sim 2.45$ and $z\sim 2.4$. We will provide a more detailed analysis of the various structures in the reconstructed density field in a subsequent paper.

\section{Conclusions and Discussion}
\label{sec:conc}
In this work we presented for the first time a comprehensive multi-survey reconstruction effort of the primordial and evolved density fields with the \texttt{COSMIC BIRTH} algorithm performed in the COSMOS field during the epoch of Cosmic Noon ($1.4\leq z \leq 3.6$). To our knowledge, this is the  ever attempted  large-scale structure analysis of its kind at high redshift,  probing the quasi-linear regime of gravitational structure formation, before non-linear shell-crossing started dominating the emerging cosmic web.

We combined the data from five spectroscopic galaxy surveys in the COSMOS field, which have partially overlapping footprints, and do not share a common observing strategy. Therefore, we  had to make a special effort in estimating the angular selection function for each survey, based on the the individual targeting strategies for the parent photometric catalogs, which was missing for these surveys. 

Also, we applied for the first time a multi-tracer and multi-survey likelihood formalism in Lagrangian space within a Bayesian inference framework. This allowed us to combine surveys with different selection functions, galaxy bias, number densities, and redshift ranges, connected through the underlying dark matter distribution. Although earlier works have presented reconstructions using multiple galaxy populations, they however shared a unique survey geometry. Therefore, our new method is more general and has a wider range of possible applications. Since we expect a correlation between the spatial distribution of the galaxy catalogues, a covariance term would in principle be necessary when different observations are combined within one joint likelihood analysis.
We bypass this problem by firstly mapping all tracers to Lagrangian coordinates. This allows us to assume Poisson likelihoods, since only in the homogeneous epoch before gravity coupled separated spatial regions, identical and independently distributed large-scale structure tracers can be assumed.

Despite of the large number of surveys, the \texttt{COSMIC BIRTH} code showed an efficient performance, converging within about 40 iterations, and showing low correlation lengths of about 40 Gibbs-sampling iterations. 

The resulting reconstructions reveal for the first time a holistic view on the matter density field and its primordial fluctuations jointly inferred from five spectroscopic surveys. 

We also revised the bias of star forming galaxies towards high redshifts, finding some moderate evidence for a stronger evolution than the one described by passive evolution. 

The inferred density fields have a large number of potential applications.
In particular, we have found several high density regions across the whole reconstructed volume. We successfully reconstructed a number of observationally known proto-cluster regions previously reported by \citet[][]{Cucciati2014, Chiang2015,Diener2015,Casey2015,Lee2016,Wang2016, ZFIRE2016,Darvish2020}. However, these previous proto-cluster studies were typically performed using individual galaxy surveys directly on the observed galaxy positions in redshift-space. Major improvements have been done combining two surveys and statistically sampling the redshifts in two-dimensional slices including a Voronoi tessellation to estimate the galaxy density field \citep{Cucciati2018}. Still, the mass estimates are prone to projection effects due to their peculiar velocities \citep{Kaiser1984,Kaiser1987}, and also assuming a velocity dispersion measure which is not valid for non-virialised objects at high redshifts. In a subsequent paper, we will carry out a more detailed analysis of these structures. Since we have reconstructed the initial conditions of the COSMOS volume, we will be able to run constrained $N$-body simulations based on the inferred initial conditions, enabling us to study the full cosmic evolution of the proto-clusters in detail (Ata et al. in prep.).
This will, for example, permit us to direct model the late-time properties of the galaxy clusters that will coalesce from the proto-clusters observed in the COSMOS
surveys. Furthermore, the Bayesian formalism will permit us to
directly quantify the uncertainty of late-time cluster properties.

In addition to studying the large-scale structure evolution, there are many potential applications for the contemporary density field derived in COSMOS. Our density map could be used to directly address the question of galaxy evolution in the context of environment \citep{Nuza2014},
without having to use contrived statistics (e.g.\ counts in cylinders, $N$-nearest neighbours) to define the galaxy environment \citep{Cooper2008,Koyama2013,Kawinwanichakij2017,Muldrew2018,Ji2018}.
More remarkably, by treating the observed COSMOS galaxies as tracer particles in the constrained $N$-body simulations based on these reconstruction, we will also be able to track them to their eventual $z=0$ environments and thus link them directly with well-studied trends in the Local Universe \citep[e.g.,][]{kauffmann:2004}.

Another clear application is to compare the dark matter reconstructions from galaxy tracers with reconstructions obtained from hydrogen Lyman-$\alpha$ forest tomography observations in the CLAMATO survey (Lee et al. in prep.). This will allow to directly test the fluctuating Gunn-Peterson approximation which posits a monotonic relationship between Lyman-$\alpha$ forest absorption  and the underlying density field. While the CLAMATO data is reconstructed moderate-resolution,
low-S/N LBG spectra, there also exist several high-resolution absorption spectra  that have been observed of bright quasars in the field. On these high-resolution  spectra, we will aim also to carry out an analysis of the line widths and column densities as a function of the underlying matter density, which will shed light
into a recent study which suggested that the thermal properties of the intergalactic medium varies between low-density and higher-density regions \citep{rorai:2018}.

The method applied in this study is also interesting for the Galaxy Evolution component of the planned Subaru PFS Subaru Strategy Program, which  will target $\sim 12-15$ deg of deep spectroscopy over three continuous fields, i.e.\ an order-of-magnitude larger area than COSMOS. The galaxy number densities are also very comparable, or better, than those reconstructed in this paper. For the lower-redshift NIR-selected sample at $z\sim 1-1.5$, the PFS will obtain spectra for galaxies at a number density of $n \approx 3 \times 10^{-3} \,h^{3} \,\mathrm{Mpc}^{-3}$, slightly better than the FMOS-COSMOS sample which has $n \sim  2\times 10^{-3} \,h^{3} \,\mathrm{Mpc}^{-3}$ at $z\sim 1.7$; This should enable direct cosmic web analysis from the galaxy reconstructions alone. Closer to the peak of Cosmic Noon at $z\sim 2-3$, an optically-selected LBG sample is planned to yield number densities equivalent to VUDS or zCOSMOS-deep ($n \sim  3\times 10^{-4} \,h^{-3} \,\mathrm{Mpc}^{-3}$). While this is less than the combined survey sample used in this paper, it should still be sufficient to identify and characterise proto-clusters within the volume.

As wide-field spectroscopic surveys probe ever deeper into cosmic history, the utility of density reconstruction techniques will become increasingly important to fully exploit the rich scientific possibilities that will open up. 

\section*{Acknowledgements}
This work was supported by the JSPS KAKENHI Grant Number JP18H05868.
The authors thank Bahram Mobasher and Karl Glazebrook for their help with the MOSDEF and ZFIRE surveys, respectively. 
MA thanks Jiaxin Han, Ben Granett and Masahiro Takada for helpful discussions and the hospitality at the IAC.
FSK thanks for the support from the grants RYC2015-18693, SEV-2015-0548 and AYA2017-89891-P.
KGL acknowledges support from JSPS KAKENHI Grant Number JP19K14755.
We commemorate our co-author Olivier Le F{\`e}vre, who sadly passed away while this paper was under review.

\section*{Data Availability Statement}
The data underlying this article will be shared on reasonable request to the corresponding author.


\bibliographystyle{mnras}
\bibliography{bib} 



\appendix

\section{Impact of an Inaccurate Bias Treatment}
\label{app:bias_wrong}

Even though the data region within the reconstructed mesh grid only occupies a small fraction, wrong galaxy bias estimates can have a strong impact on the accuracy of the reconstructions, as we demonstrate here. 
In particular, we perform a \texttt{COSMIC BIRTH} run on a sub-volume covering the redshift range of $1.8<z<2.2$ with the data described in Section \ref{sec:surveys}.  We assume a large scale bias  $\sim 60\%$ systematically higher than the one from our regular runs, as presented in Figure \ref{fig:bias_compare} and discussed in Section \ref{sec:bias}. This bias  exceeds the mean   observational measurements in that redshift range by 2 $\sigma$, and therefore represents a conservative upper bias  limit  \citep[see the measurement by][extrapolated to higher redshifts with passive evolution in Figure \ref{fig:bias_compare}]{Kashino2017}.

\begin{figure}
    \centering
    \includegraphics[width=0.48\textwidth]{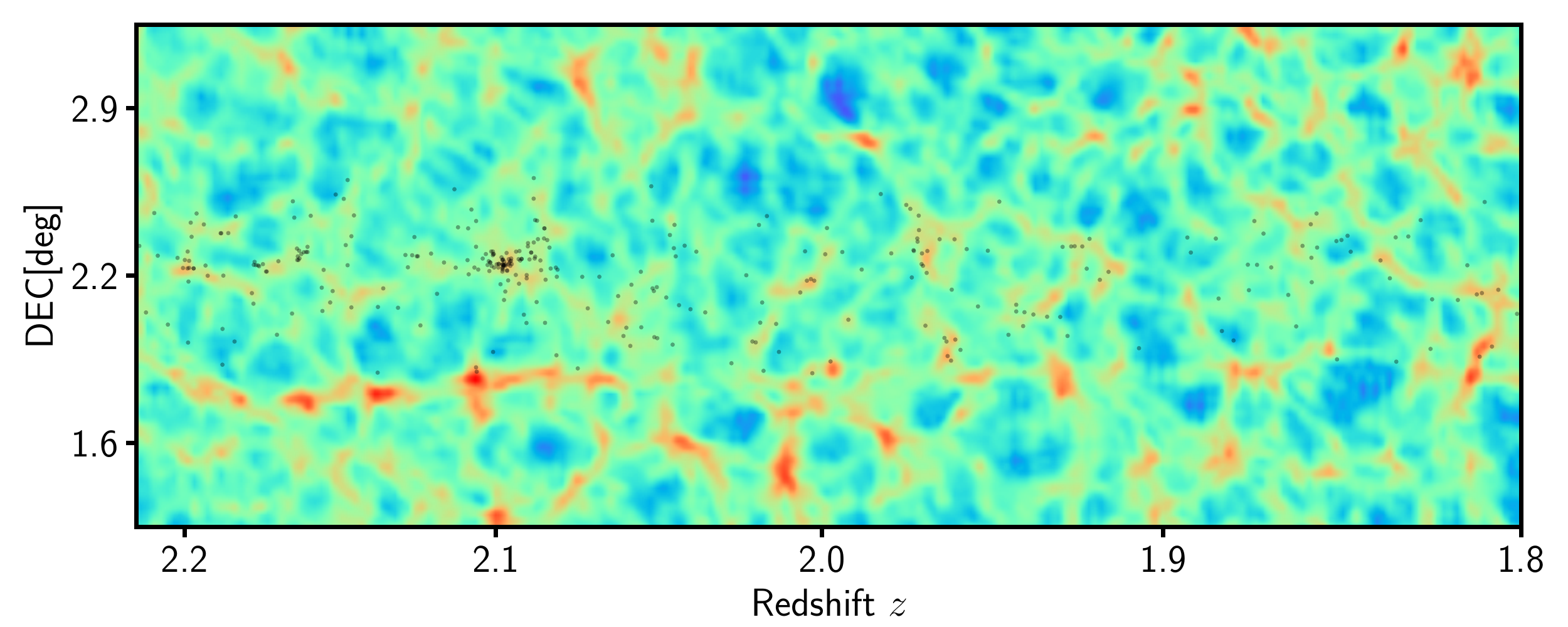}
    \hspace*{-0.5cm}
    \includegraphics[width=0.49\textwidth]{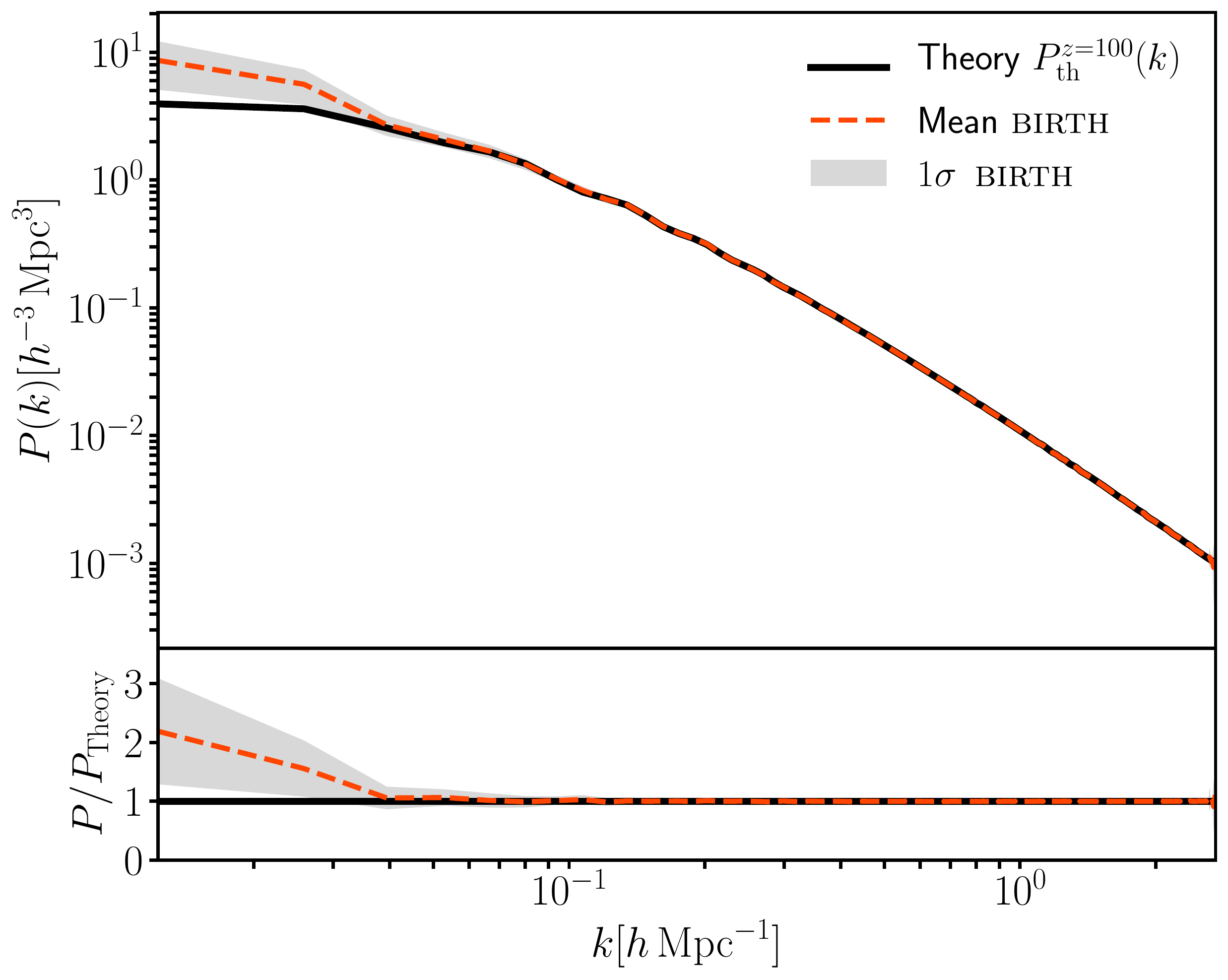}
    \caption{Results from \texttt{COSMIC BIRTH} runs with a bias exceeding the mean measurement given by the literature by 2 $\sigma$ in the redshift range $1.8<z<2.2$. {\bf Top}: Slice plot of the light-cone density field $\delta(\mbi s)$ with the galaxy positions plotted on top represented by  black dots.
    {\bf Bottom}: Power spectra of the corresponding initial density fields $\delta(\mbi q)$. We show the mean (blue dashed line) and the standard deviation (grey band)  for 1000 HMC realizations, including the ratio with the theoretical $\Lambda$CDM power spectrum in the panel below.}
\label{fig:rec_wrong}
\end{figure}

The results for this test are represented in Figure \ref{fig:rec_wrong}. On the top we show the inferred density field $\delta(\mbi s)$ with the galaxy positions plotted on top represented by black dots. Although the reconstructed structures match the galaxy positions, a deeper inspection shows that the data region around $\mathrm{DEC}=2.2$ is under-weighted, due to the overestimated bias value and artificially high density structures at the edge of the data region are formed.
In the bottom plot we show the mean power spectrum (red dashed) and the $1\sigma$ standard deviation (grey band) of the inferred initial density fields for 1000 samples after convergence including the ratio of the inferred power spectra with the theoretical $\Lambda$CDM prediction below. 
Considering the low volume filling fraction, unbiased power spectra with the theoretical one are expected. However, a clear excess of power can be seen at the lowest modes, excluding the proposed bias values in this test.
A thorough analysis of the bias, will be presented in a forthcoming paper.

\section{Different individual cluster reconstructions}
We focus on the prominent $z\sim2.1$ (mainly seen in ZFIRE) proto-cluster and analyze the variance within the chain by looking at the result of 10 individual HMC reconstructions, that we choose randomly.
We find on average a $\delta_{\rm sum} = 841 \pm 121$ when we sum all cells in a $10 \mperh$ cubic mask around the centre of the proto-cluster.  
\label{app:cluster}
 \begin{figure*}
    \includegraphics[width=0.39\textwidth]{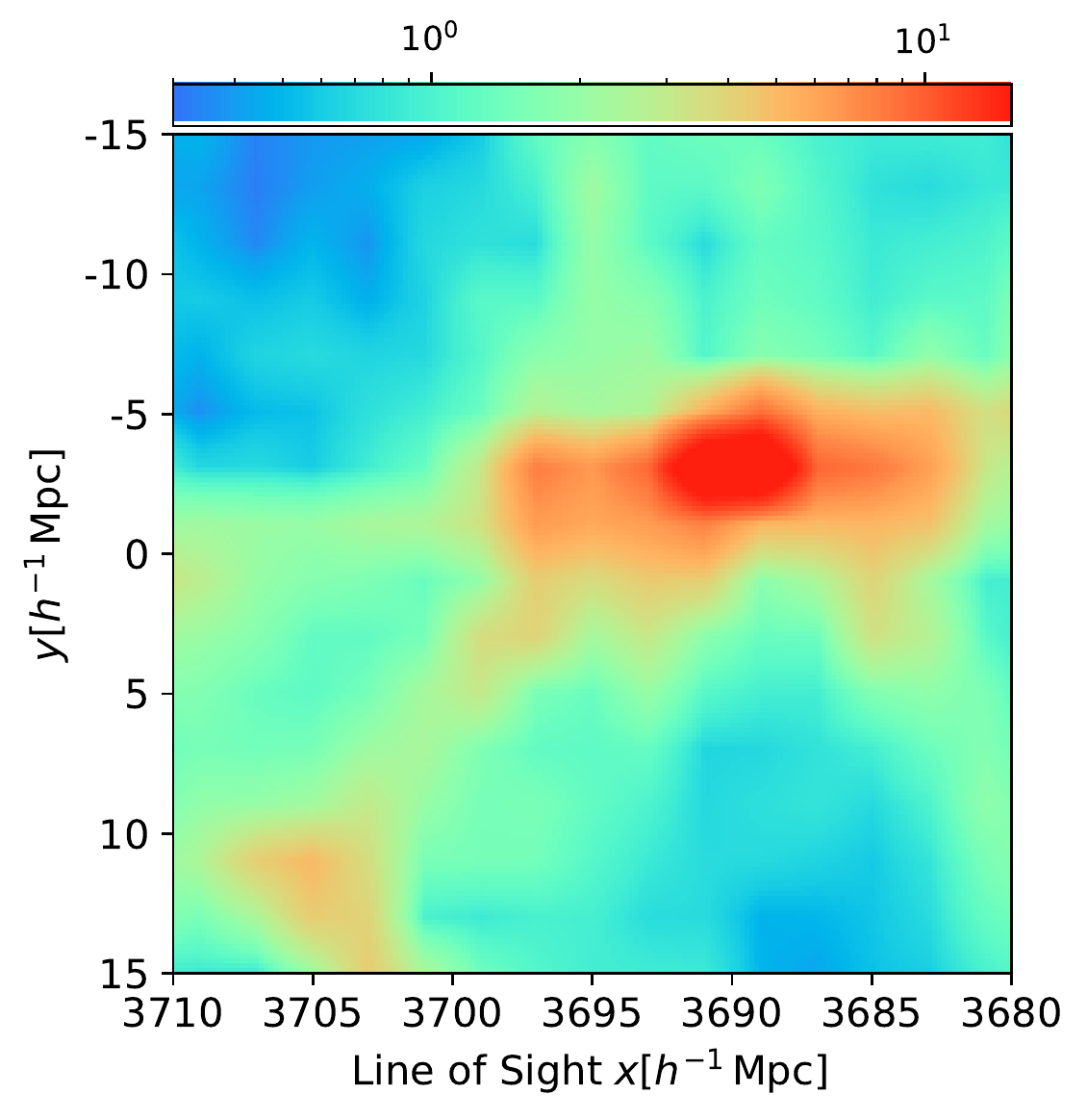}
    \includegraphics[width=.39\textwidth]{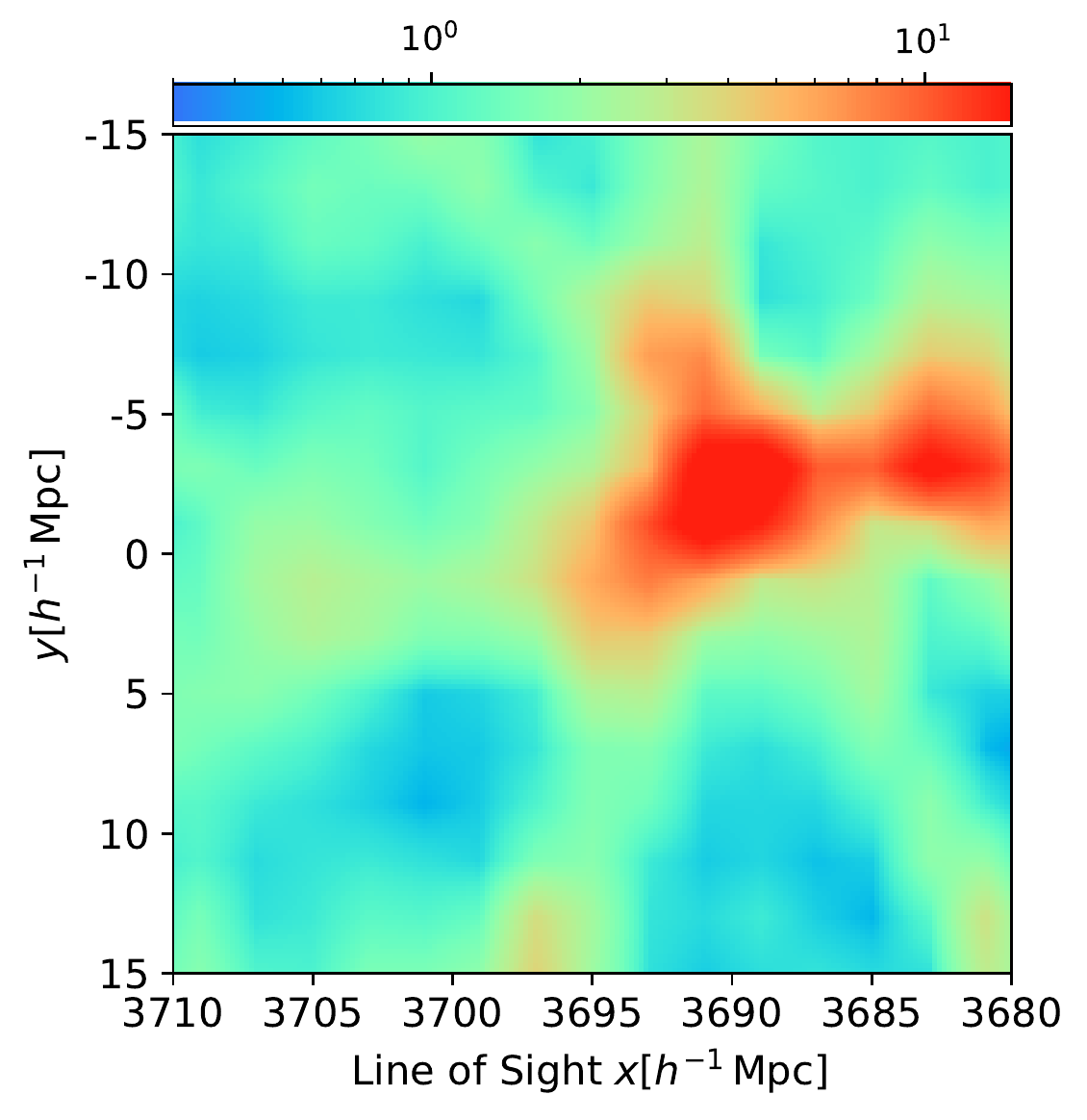}
    \includegraphics[width=.39\textwidth]{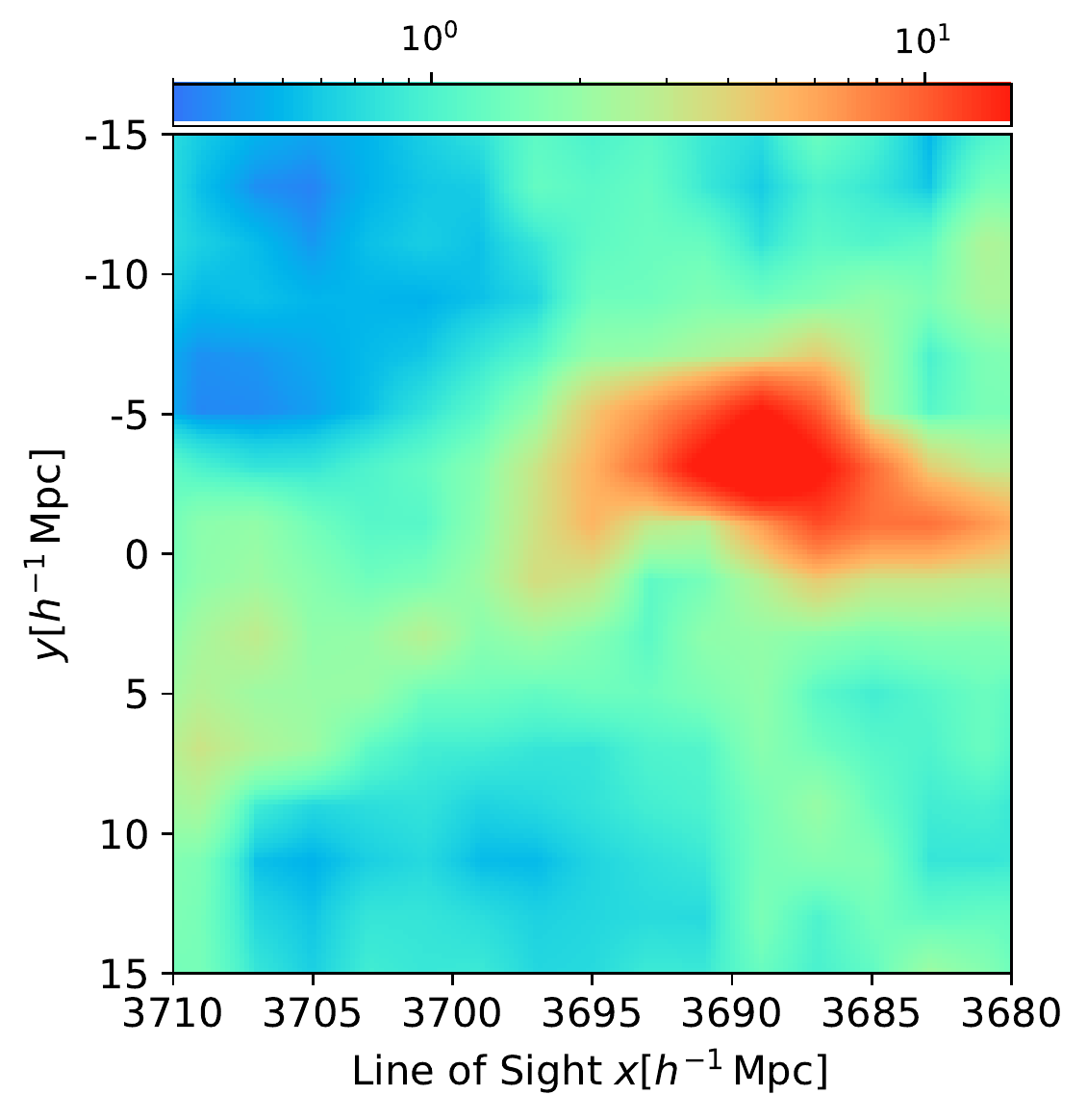}
    \includegraphics[width=.39\textwidth]{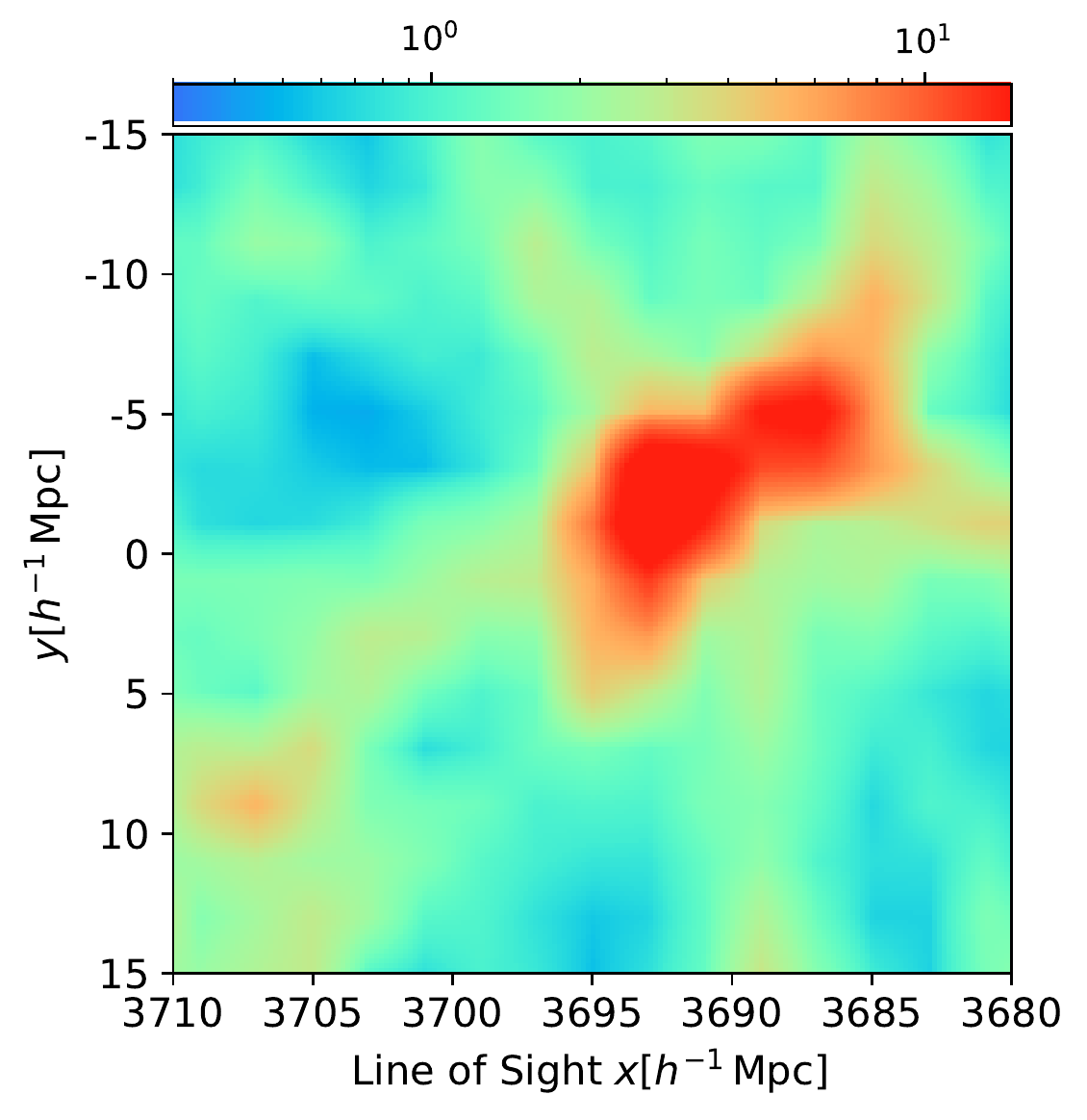}
    \includegraphics[width=.39\textwidth]{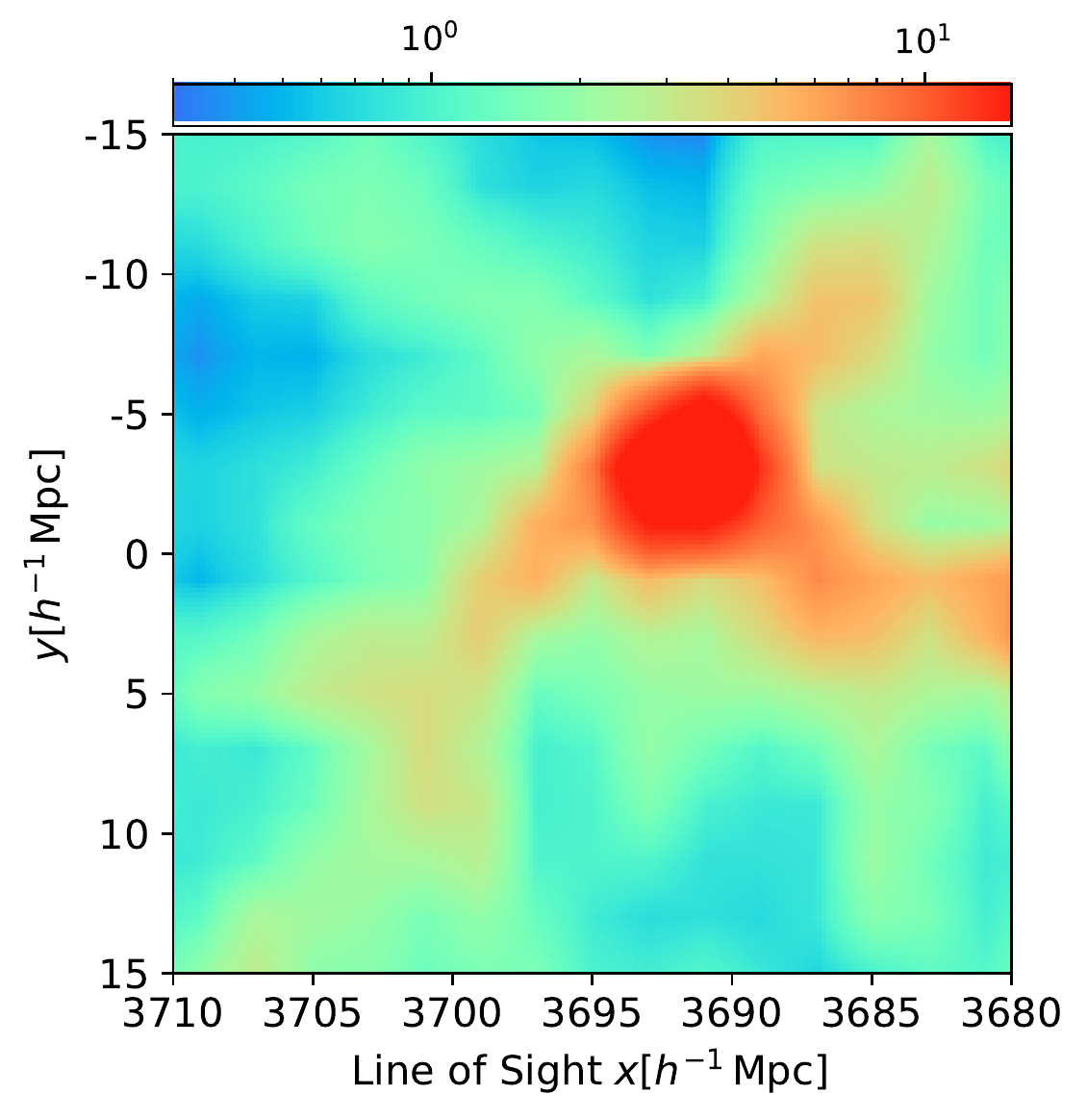}
    \includegraphics[width=.39\textwidth]{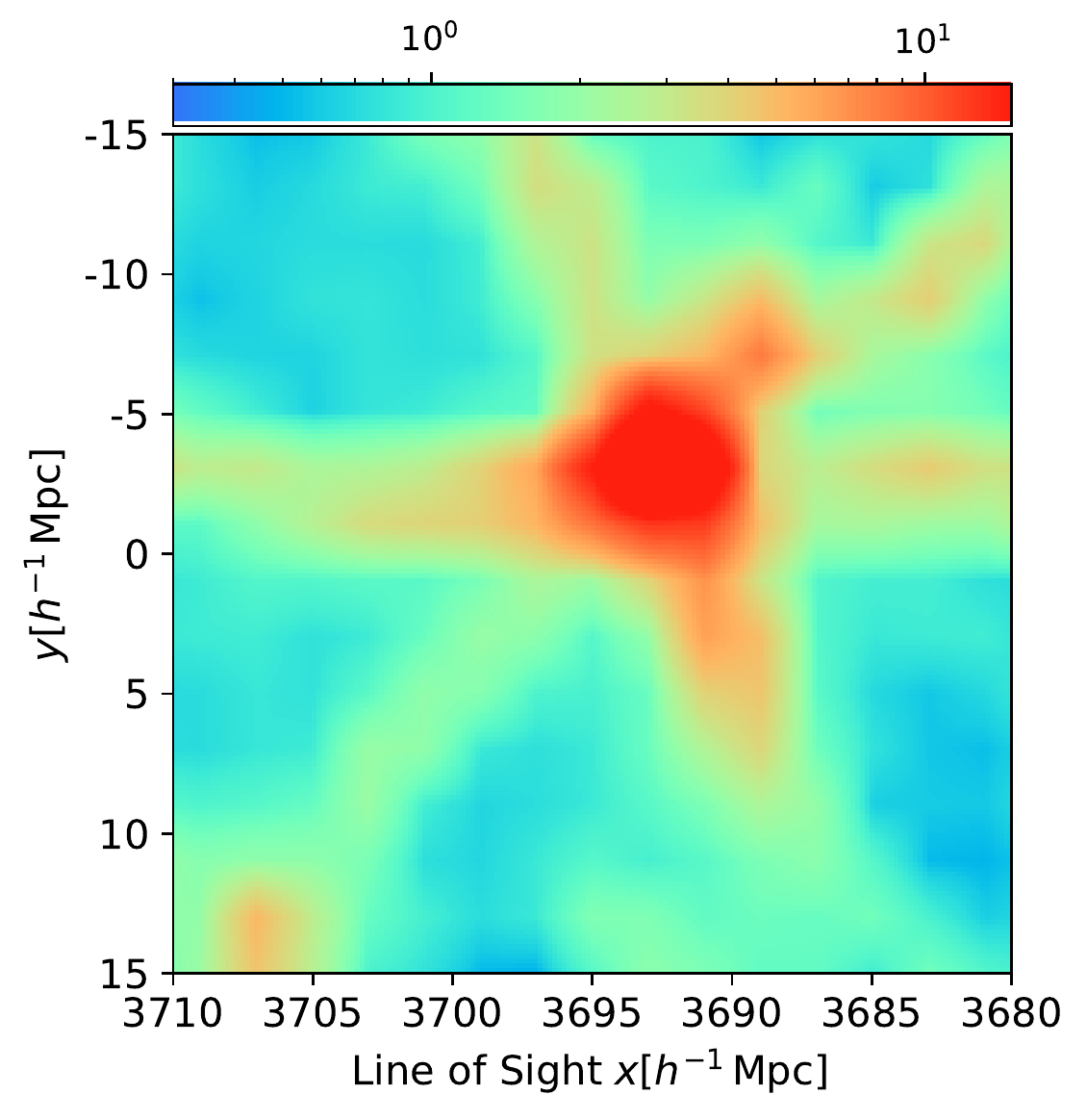}
    \caption{Cluster reconstruction among different, randomly chosen HMC realizations}
    \label{fig:cluster}
\end{figure*}


\bsp	
\label{lastpage}
\end{document}